\documentclass[titlepage, 12pt]{utarticle}
\usepackage{epsfig}
\usepackage{graphicx}
\usepackage{hyperref} %uncomment for HyperTeX links
\usepackage{graphics}
\usepackage{amsmath,amssymb,amscd}
\usepackage{color}
\usepackage{latexsym}
\usepackage{ifthen}
\usepackage{bm,longtable,enumerate}

\numberwithin{equation}{section}

\newcommand\eps{\epsilon}

\begin{document}
\preprint{{\tt arXiv:1006.2363}\\
}

\title{Pion and Vector Meson Form Factors in the \vskip0.5cm Kuperstein-Sonnenschein holographic model}

\author{C.A. Ballon Bayona,
	 \address{
	  Centro Brasileiro de Pesquisas F{\'i}sicas,\\
          Rua Dr. Xavier Sigaud 150, Urca, 22290-180 Rio de Janeiro, RJ, Brasil\\
	  {\tt ballon@cbpf.br}
        }
	H. Boschi-Filho, Matthias Ihl and M.A.C. Torres
	 \address{
	 Instituto de F{\'i}sica,\\
	 Universidade Federal do Rio de Janeiro,\\
         Caixa Postal 68528,\\
	 21941-972 Rio de Janeiro, RJ, Brasil\\
        {\tt boschi,msihl,mtorres@if.ufrj.br}}
	 }

\Abstract{We study phenomenological aspects of the holographic model of chiral symmetry breaking recently introduced by Kuperstein and Sonnenschein (KS). As a first step, we calculate the spectrum of vector and axial-vector mesons in the KS model. We numerically compute various coupling constants of the mesons and pions. Our analysis indicates that vector meson dominance is realized in this model.
The pion, vector meson and axial-vector meson form factors are obtained and studied in detail. We find good agreement with QCD results. In particular, the pion form factor closely matches available experimental data.}

\maketitle
\tableofcontents
\newpage

\section{Introduction}
Holographic models based on the gauge/gravity duality (descending from the AdS/CFT correspondence, originally introduced by Maldacena, see e.g., \cite{Maldacena:1997re, Witten:1998qj}) have been studied and improved significantly over the last decade.
Though still unable to provide a gravity dual for QCD proper, some of the signature features of QCD have been incorporated into semirealistic models.
An important improvement was made within the model by Sakai and Sugimoto \cite{Sakai:2004cn, Sakai:2005yt}, who implemented a geometrical mechanism for chiral symmetry breaking into Witten's model \cite{Witten:1998zw}. The construction is based on the following strategy. The starting point is the 
near horizon limit of a geometry generated by a large number $N_c$ of color $D$-branes. To include fundamental quarks into the dual gauge theory one has to add a stack of $N_f$ flavor $D$-branes to this background. Strings stretching between the color and flavor branes naturally transform in the fundamental representation $U(N_c) \times U(N_f)$. In case $N_f \ll N_c$, the backreaction of the flavor branes on the background can be safely 
neglected, i.e., one works in the probe brane approximation \cite{Karch:2002sh}. This corresponds to the so-called quenched approximation of (large $N_c$ massless) lattice QCD, in which the backreaction of the flavors on the colors is neglected completely while the color dynamics and its effect on the flavors are accounted for.\\
In order to have a geometric mechanism for chiral symmetry breaking one needs to add another stack of $N_f$ anti-$D$-branes associated in the dual gauge theory to anti-quarks in the fundamental representation of $U(N_f)$. This has to be accomplished in such a way that in the UV the two stacks are 
separated from each other, resulting in the dual gauge theory enjoying the full $U_L(N_f) \times U_R(N_f)$ flavor symmetry. In the IR however, if the two
stacks of branes and anti-branes merge smoothly into a single stack, only a diagonal subgroup $U_D(N_f)$ of the full flavor group survives and chiral symmetry is broken.\\
In the Sakai-Sugimoto model, this was beautifully realized by introducing $N_f$ $D8$- and $\overline{D8}$-branes into the near extremal geometry generated by a stack of $N_c$ $D4$-branes in type IIA superstring theory. In order to obtain a $3+1$ dimensional gauge theory in the IR, the $x_4$-direction is compactified on a circle. 
The two-dimensional submanifold of the background in the $(x_4,r)$-directions turns out to be cigar-shaped, leading to a U-shaped profile of the trajectories of the $D8$-$\overline{D8}$-branes in these directions, thus constituting a geometric realization of chiral symmetry breaking as described above.\\
The Sakai-Sugimoto model has been applied to a large number of different aspects in hadronic physics. For a review of mesons in this and other holographic models, see \cite{Erdmenger:2007cm} and references therein. Form factors of vector and axial-vector mesons and pions in the Sakai-Sugimoto and other holographic models have been obtained in, e.g., \cite{BallonBayona:2009ar,Hong:2004sa,Grigoryan:2007vg,Brodsky:2007hb}.\\
Although the Sakai-Sugimoto model was the first example of a stable geometric model of chiral symmetry breaking with completely broken supersymmetry, it suffers from shortcomings inherited from Witten's original model. Most notably, the model is only well-defined as a $3+1$-dimensional field theory 
in the IR. As one goes to higher energies, KK modes become important at masses of the order of the inverse radius of the $S^1$. At even higher energies,
the dilaton (or string coupling) starts to blow up and one has to S-dualize. The UV completion of the theory is in fact the six-dimensional $A_{N_c-1}$ $(2,0)$ superconformal theory \cite{Aharony:2006da,Strominger:1995ac}.
To overcome some of these problems, Kuperstein and Sonnenschein \cite{Kuperstein:2008cq} managed to incorporate stacks of flavor $D7$- and $\overline{D7}$-branes into a background with constant dilaton, namely the Klebanov-Witten model \cite{Klebanov:1998hh}, which is based on the singular conifold geometry. This embedding was engineered such that it breaks the remaining supersymmetry of the background completely and the two branches join smoothly in the IR.\footnote{A lot of work has been done studying supersymmetric embeddings into the Klebanov-Witten and Klebanov-Strassler models (e.g., \cite{Ouyang:2003df, Kuperstein:2004hy, Levi:2005hh}). However, it turns out that these embeddings only allow a single branch of branes and thus they are not useful as models of chiral symmetry beaking.}\\ 
In this paper we will focus on studying phenomenological aspects of the KS model \cite{Kuperstein:2008cq}, such as the spectrum of vector and axial-vector mesons, their mutual couplings and couplings to the pion as well as form factors of pions and (axial-) vector mesons. Our analysis indicates that vector meson dominance is realized in the KS model which means that the electromagnetic interaction of hadrons is mediated exclusively by vector mesons. The mass spectrum of (axial-) vector mesons was first obtained in \cite{Kuperstein:2008cq}, and our findings agree with the mass spectrum presented in that article.
Many of the results presented below are quite similar to the results in the Sakai-Sugimoto $D4-D8$ brane model \cite{Sakai:2004cn,Sakai:2005yt,BallonBayona:2009ar}. This is due to the similar shape of the $D7$-brane embedding. For the pion form factor, our results in fact reproduce the experimental data slightly better than the Sakai-Sugimoto model, as we will demonstrate below. Besides obvious numerical differences, one crucial difference is that, in the KS model, instead of a Kaluza-Klein mass scale $M_{\text{KK}}$, there is a mass scale $M_{\ast}$ depending on the parameter $r_0$, the minimal distance of the $D7$-branes to the singularity at $r=0$, which also parametrizes a one-parameter family of solutions to the classical equations of motion of the $D7$-brane embedding.\\
The paper is organized as follows: \\
In section \ref{sec:KS}, we give a brief review of the Kuperstein-Sonnenschein model, followed by a discussion of vector and axial-vector mesons in 
section \ref{sec:mesons}. Section \ref{sec:pion} contains a discussion of the pions and vector meson dominance. In section \ref{sec:form}, we present our results for the various form factors. We formulate our conclusions in section \ref{sec:concl}. 

\section{Review of the Kuperstein-Sonnenschein model}\label{sec:KS}

\subsection{D3-brane background}
The KS model is based on the $D3$-brane background with a conical singularity in type IIB superstring theory first studied by Klebanov and Witten \cite{Klebanov:1998hh}. This is an extension of the original AdS/CFT correspondence by studying strings on $AdS_5 \times X^5$, where $X^5$ no longer 
is the 5-sphere but a five-dimensional Sasaki-Einstein space, namely the coset space $T^{1,1}= (SU(2) \times SU(2))/U(1)$ with Einstein metric
\begin{equation}
ds^2_{T^{1,1}} = \frac{1}{9} \left(d \psi + \sum_{i=1}^2 \cos \theta_i d \phi_i \right)^2 + \frac{1}{6} \sum_{i=1}^2 (d \theta_i^2 + \sin^2 \theta_i d \phi_i^2).
\end{equation}
Before taking the near-horizon limit, the six dimensional internal space transverse to the $D3$-branes is a Calabi-Yau manifold $Y_6$ with a singularity, called a conifold, which is a cone over the base manifold $T^{1,1}$ (topologically, $T^{1,1}$ is $S^2 \times S^ 3$). Namely,
\begin{equation}
ds^2_{(10)}= \frac{r^2}{R^2} \eta_{\mu \nu} dx^{\mu} dx^{\nu} + \frac{R^2}{r^2} ds^2_{(6)},
\end{equation}
where 
\begin{equation}
ds^2_{(6)}= \left( dr^2 + r^2 ds^2_{T^{1,1}} \right), 
\end{equation}
and the $AdS_5$ radius is $R^4 = \frac{27}{4} \pi g_s N_c l_s^4$.\\
The field theory dual resulting from a stack of $N_c$ $D3$-branes placed at the conifold singularity is a ${\mathcal N}=1$ superconformal
field theory with gauge group $SU(N_c) \times SU(N_c)$ and global symmetry $(SU(2) \times SU(2)) \times U(1)$. It contains four chiral superfields, 
$A_i, i=1,2$, a doublet of the first $SU(2)$ factor transforming as $(N_c,\overline{N_c})$ and $B_j, j=1,2$, a doublet of the second $SU(2)$ factor transforming as $(\overline{N_c},N_c)$. The theory has a marginal superpotential ${\mathcal W}= \eps^{il} \eps^{jm} A_i B_j A_l B_m$, and the R-charge of
all chiral superfields is $\frac{1}{2}$.

\subsection{D7-brane profiles} 
One of the achievements of \cite{Kuperstein:2008cq} was to study a $D7$-brane configuration spanning the space-time coordinates $x^{\mu}$, the radial
coordinate $r$ and the $S^3$ parametrized by the one-forms $f_i$.
The one-forms $f_i, i=1,2,3$ are defined as (for details, cf.~\cite{Minasian:1999tt,Gimon:2002nr,Evslin:2007ux,Krishnan:2008gx}),
\begin{equation}
 \left( \begin{array}{c} f_1 \\ f_2 \\ f_3 \end{array} \right)= \left(\begin{array}{ccc} 0 & \cos \theta & - \sin \theta \\ 1 & 0 & 0 \\
0 & \sin \theta & \cos \theta \end{array}\right) \left(\begin{array}{ccc} - \sin \phi & - \cos \phi & 0 \\ -\cos \phi & \sin \phi & 0 \\
0 & 0 & 1 \end{array}\right)  \left( \begin{array}{c} w_1' \\ w_2' \\ w_3' \end{array} \right)
\end{equation}
where the $SU(2)$ left-invariant Maurer-Cartan forms $w_i', i=1,2,3$ are determined by the following condition:
\begin{equation}
X^{\dag} d X =  \frac{i}{2}\sum_{i=1}^3 \sigma_i w_i'.
\end{equation}
The explicit form of the one-forms $w_i'$ and the definition of the matrix $X$ can be found, e.g., in \cite{Krishnan:2008gx}.
It should be noted that the structure involving the $SO(3)$ matrices determines the fibration of the $S^3$ over the $S^2$.
This leaves two transversal coordinates, namely those on the $S^2$, $\theta$ and $\phi$. We will assume that $\theta$ and $\phi$ do not depend on the $S^3$ coordinates (note that our ansatz preserves one of the $SU(2)$ factors of the global conifold symmetry). This can be justified by expanding the action around the solution and verifying that contributions from non-trivial $S^3$ modes only appear at second order in the fluctuations. Therefore the classical solution will only depend on the radial coordinate, $\theta = \theta (r)$, $\phi =\phi (r)$.
Plugging this ansatz into the DBI action for $D7$-branes yields the following Lagrangian (for details, see the discussion in \cite{Kuperstein:2008cq}):
\begin{equation}
{\mathcal L} \propto r^3 \left( 1 + \frac{r^2}{6} \left[ \left(\frac{\partial{\theta}}{\partial r}\right)^2 + \sin^2 \theta \left(\frac{\partial{\phi}}{\partial r}\right)^2 \right] \right). 
\end{equation}
We can use the $SU(2)$ invariance of the Lagrangian to set $\theta=\pi/2$, i.e., restrict the motion to the equator of the $S^2$. Then the solution of the field equations is given by
\begin{equation}\label{eq:D7sol}
\cos \left( \frac{4}{\sqrt{6}}\phi(r)\right)=\left(\frac{r_0}{r}\right)^4. 
\end{equation}
The trajectory $\phi(r)$ has two branches, namely $\phi \in [-\pi/2,0]$ and $\phi \in [0,\pi/2]$. In the extremal case $r_0=0$, there are two r-independent solutions, $\phi_{\pm} = \pm \frac{\sqrt{6}}{8} \pi$. For $r_0 \neq 0$, the solution $\phi(r)$ starts at $\phi(r=r_0)=0$ and approaches the asymptotic values $\phi_{\pm}$ as $r \rightarrow \infty$.\\
In conclusion, \cite{Kuperstein:2008cq} found a one-parameter family of solutions parametrized by $r_0$ sharing the same boundary values $\phi_{\pm}$. For 
$r_0 >0$ the conformal symmetry of the underlying Klebanov-Witten background is broken, since there is no $AdS_5$ factor in the induced metric. It is worth noting, however, that in the limit $r \rightarrow \infty$, one recovers the $AdS_5$ metric and conformal invariance is restored. As a consequence of this asymptotic behavior, we expect that the KS embedding reproduces QCD results better than the Sakai-Sugimoto model in the UV regime.\\
Introducing the coordinates $y= r^4\cos \left( \frac{4}{\sqrt{6}}\phi \right)$ and $z =r^4\sin \left( \frac{4}{\sqrt{6}}\phi \right)$ and noting that
$y=r_0^4$ is constant along the classical configuration solving (\ref{eq:D7sol}), the induced metric on the $D7$-branes can be written as
\begin{equation}\label{eq:D7metric}
 ds_{(8)}^2 = \frac{r^2}{R^2} \eta_{\mu \nu} dx^{\mu} dx^{\nu} + R^2 \left( \frac{(z^2 + 2 r_0^8)}{16 r^{16}}dz^2 - \frac{\sqrt{6}r_0^4}{12 r^8}dz f_1
 + \frac{1}{6} (f_1^2 + f_2^2) +\frac{f_3^2}{9}\right),
\end{equation}
where $r^8 = z^2 + r_0^8$ determines the relation $r=r(z)$. 

\section{Vector and axial-vector mesons}\label{sec:mesons}
The strategy for calculating vector meson spectra utilized here will closely follow \cite{BallonBayona:2009ar}. According to the gauge/gravity correpondence, mesons of the dual gauge theory arise as modes of open strings stretching between the $D7$- and $\overline{D7}$-branes. In particular, vector and axial-vector mesons correspond to fluctuations of the $U(N_f)$ gauge fields living on the $D7$-branes. To first order, these fluctuations
are captured by the action
\begin{equation}\label{eq:D7action}
S_{D7} = -(\pi \alpha')^2 \mu_7 \int d^4x dz d^3\Omega e^{-\Phi} \sqrt{-g_{(8)}} \text{tr} (F_{MN} F^{MN}) + S_{\text{CS}},
\end{equation}
where the first term is the Maxwell term of the non-abelian Dirac-Born-Infeld action\footnote{We neglect higher order contributions to the DBI action which are suppressed by higher powers of $\alpha'$.} of the $D7$ branes and the second term is the Chern-Simons term, which 
is irrelevant for the topic at hand. The non-abelian field strength $F_{MN} =\partial_M A_N -\partial_N A_M + [A_M, A_N]$ contains the gauge field components $A_M= (A_{\mu}, A_z, A_{\alpha})$, $\mu=0,1,2,3$, $\alpha=5,6,7$ and $g_{(8)}$ is the determinant of the induced metric on the $D7$-branes, given by (\ref{eq:D7metric}). In what follows, we will set $A_{\alpha} =0$ and assume that the other gauge field components are independent of the $S^3$ coordinates.\\\\
\noindent
{\bf Five-dimensional effective action.}
Integrating the $D7$-brane action (\ref{eq:D7action}) over the $S^3$ coordinates leads to a five-dimensional effective action for the fluctuations of the gauge fields,
\begin{equation}\label{eq:5daction}
S_{5d, \text{eff}} = -\kappa \int d^4x \int d\tilde{z} \; \text{tr} \left[\frac{1}{2} C(\tilde{z}) \eta^{\mu \lambda} \eta^{\nu \rho} F_{\lambda \rho} F_{\mu \nu} + 
M_{\ast}^2 D(\tilde{z}) \eta^{\mu \nu} F_{\mu \tilde{z}} F_{\nu \tilde{z}} \right],   
\end{equation}
where $\kappa= 2(\pi\alpha')^2 \mu_7 g_s^{-1} \frac{R^4}{72} \text{Vol}S^3 = \frac{3 N_c}{1024 \pi^2}$. Here we have defined the dimensionless variable $\tilde{z} = \frac{z}{r_0^4}$ and an $r_0$-dependent mass scale $M_{\ast}^2(r_0) = \frac{16r_0^2}{R^4}$. The functions  
\begin{equation}
C(\tilde{z}) =(\tilde{z}^2 + 1)^{-1/2} \; \text{and} \; D(\tilde{z}) = (\tilde{z}^2 + 1)^{3/4},
\end{equation}
are rescaled, dimensionless variants of the functions $C(z)$ and $D(z)$ introduced in \cite{Kuperstein:2008cq}, eq.~(4.14).
It turns out to be convenient to work in the $A_{\tilde{z}}=0$ gauge in which $A_{\mu}$ can be expanded in the following way:
\begin{equation}\label{eq:Amuexp}
 A_{\mu}(x,\tilde{z})= \hat{\mathcal{V}}_{\mu}(x)+ \hat{\mathcal{A}}_{\mu}(x) \psi_0(\tilde{z}) + \sum_{n=1}^{\infty} v_{\mu}^{(n)}(x) \psi_{2n-1}(\tilde{z})+ \sum_{n=1}^{\infty} a_{\mu}^{(n)}(x) \psi_{2n}(\tilde{z}),
\end{equation}
where\footnote{The normalization constant is obtained by demanding 
$$ \lim_{\tilde{z} \rightarrow +\infty} A_{\mu} (x^{\mu},\tilde{z}) = A_{L\mu}^{\xi_+} (x^{\mu}), \quad \lim_{\tilde{z} \rightarrow -\infty} A_{\mu} (x^{\mu},\tilde{z}) = A_{R\mu}^{\xi_-} (x^{\mu}).$$
In the $A_{\tilde{z}}=0$ gauge, we can write
$$  A_{\mu}(x^{\mu},\tilde{z})= A_{L\mu}^{\xi_+} (x^{\mu}) \psi_+ (\tilde{z}) + A_{R\mu}^{\xi_-} (x^{\mu}) \psi_-(\tilde{z}) + \sum_{n=1}^{\infty} v_{\mu}^{(n)}(x) \psi_{2n-1}(\tilde{z})+ \sum_{n=1}^{\infty} a_{\mu}^{(n)}(x) \psi_{2n}(\tilde{z}), $$
with $\psi_{\pm}(\tilde{z}) = \frac{1}{2} (1 \pm \psi_0 (\tilde{z}))$. The normalization constant now follows from imposing
$$\lim_{\tilde{z} \rightarrow \pm \infty} \psi_0 (\tilde{z}) = \pm 1.$$
Note that $\partial_{\tilde{z}} \psi_{\pm} =\pm \frac{1}{2} \partial_{\tilde{z}} \psi_0 (\tilde{z}) \propto D(\tilde{z})^{-1}$.\label{fn:repeat}} $\psi_0(\tilde{z})= \frac{2 \Gamma (3/4)}{\sqrt{\pi}\Gamma (1/4)}\tilde{z}\; {{}_2{F}_1}\left(\frac{1}{2}, \frac{3}{4}, \frac{3}{2}, -\tilde{z}^2 \right)$, and the fields $v_{\mu}^{(n)}$ and $a_{\mu}^{(n)}$ represent vector and axial-vector mesons, respectively. The pion field $\Pi(x)$ appears in the expansion of 
\begin{eqnarray}
 \hat{\mathcal{V}}_{\mu}(x) &=& \frac{1}{2}\left( A_{L\mu}^{\xi_+}(x) + A_{R\mu}^{\xi_-}(x)\right) \\\nonumber  
 &=&\frac{1}{2} \xi_+ \left( A_{L \mu} (x) + \partial_{\mu} \right)\xi_+^{-1} + \frac{1}{2} \xi_- \left( A_{R \mu} (x) + \partial_{\mu} \right)\xi_-^{-1},\\
 \hat{\mathcal{A}}_{\mu}(x) &=& \frac{1}{2}\left( A_{L\mu}^{\xi_+}(x) - A_{R\mu}^{\xi_-}(x)\right) \\\nonumber
 &=& \frac{1}{2} \xi_+ \left( A_{L \mu} (x) + \partial_{\mu} \right)\xi_+^{-1} - \frac{1}{2} \xi_- \left( A_{R \mu} (x) + \partial_{\mu} \right)\xi_-^{-1},
\end{eqnarray}
where $A_{L\mu}(x)$ and $A_{R\mu}(x)$ are external gauge fields and we have defined $ \xi_+^{-1} = \xi_- := e^{i \frac{\Pi(x)}{f_\pi}}$. 
The normalization conditions and equations of motion for the wave functions $\psi_n (\tilde{z})$ can be written as:
\begin{eqnarray}
\kappa \int d\tilde{z} C(\tilde{z}) \psi_m(\tilde{z}) \psi_n(\tilde{z}) &=& \delta_{m,n}, \label{eq:norm} \\
- (C(\tilde{z}))^{-1} \partial_{\tilde{z}} (D(\tilde{z}) \partial_{\tilde{z}} \psi_n (\tilde{z})) &=& \lambda_n \psi_n(\tilde{z}),\label{eq:eom}
\end{eqnarray}
with $\lambda_n := M_n^2 \frac{R^4}{16 r_0^2}= M_n^2/M_{\ast}^2$.\\\\
\noindent
{\bf Four-dimensional effective action for the mesons.}
Substituting the gauge field (\ref{eq:Amuexp}) into the action $S_{5d, \text{eff}}$, eq.~(\ref{eq:5daction}), we find the four-dimensional effective Lagrangian for the vector and axial-vector mesons (ignoring divergent terms from non-renormalizable contributions), $\mathcal{L}_{4d, \text{eff}} = \mathcal{L}_{4d, \text{div}} + \sum_{j=2}^{\infty} \mathcal{L}_{4d, \text{eff}}^{(j)}$. The kinetic part of the Lagrangian reads
\begin{eqnarray}\label{eq:Lag2}
 \mathcal{L}_{4d, \text{eff}}^{(2)} &=& \frac{1}{2} \text{tr} (\partial_{\mu} \tilde{v}_{\nu}^{(n)} - \partial_{\nu} \tilde{v}_{\mu}^{(n)})^2 
 + \frac{1}{2} \text{tr} (\partial_{\mu} \tilde{a}_{\nu}^{(n)} - \partial_{\nu} \tilde{a}_{\mu}^{(n)})^2 + \text{tr} (i \partial_{\mu} \Pi + f_{\pi} \mathcal{A}_{\mu})^2 \nonumber \\
&& + M^2_{v^n}  \text{tr} \left( \tilde{v}_{\mu}^{(n)} - \frac{g_{v^n}}{M^2_{v^n}} \mathcal{V}_{\mu}\right)^2 + M^2_{a^n}  \text{tr} \left( \tilde{a}_{\mu}^{(n)} - \frac{g_{a^n}}{M^2_{a^n}} \mathcal{A}_{\mu}\right)^2,  
\end{eqnarray}
where we have used the following redefinitions of meson fields in order to diagonalize the kinetic terms:
\begin{equation}
  \tilde{v}_{\mu}^{(n)} = v_{\mu}^{(n)}+\frac{g_{v^n}}{M^2_{v^n}}\mathcal{V}_{\mu}, \quad \tilde{a}_{\mu}^{(n)} = a_{\mu}^{(n)}+\frac{g_{a^n}}{M^2_{a^n}}\mathcal{A}_{\mu}. 
\end{equation}
Moreover, we define
\begin{eqnarray}
M_{v^n}^2 &=& \lambda_{2n-1} M_{\ast}^2, \quad M_{a^n}^2 = \lambda_{2n} M_{\ast}^2, \\
\mathcal{V}_{\mu} &=& \frac{1}{2} (A_{L\mu} + A_{R\mu}), \quad  \mathcal{A}_{\mu} = \frac{1}{2} (A_{L\mu} - A_{R\mu}).
\end{eqnarray}
The constants 
\begin{eqnarray}
g_{v^n} &=& \kappa M_{v^n}^2 \int_{-\infty}^{+\infty} d\tilde{z} C(\tilde{z}) \psi_{2n-1} (\tilde{z}), \label{eq:gv}\nonumber \\
&=& -2 \kappa M_{\ast}^2 \left( D(\tilde{z}) \partial_{\tilde{z}} \psi_{2n-1}(\tilde{z})\right)\Big{|}_{\tilde{z}\rightarrow \infty}, \\
g_{a^n} &=& \kappa M_{a^n}^2 \int_{-\infty}^{+\infty} d\tilde{z} C(\tilde{z}) \psi_{2n} (\tilde{z})\psi_0 (\tilde{z}),
\end{eqnarray}
are the couplings between a massive vector meson $\tilde{v}_{\nu}^{(n)}$ (axial-vector meson $\tilde{a}_{\nu}^{(n)}$) and an external $U(1)$ field $\mathcal{V}_{\mu}$ representing a photon (an external axial $U(1)$ field $\mathcal{A}_{\mu}$).\\
The interaction part of the effective Lagrangian governing the interactions among vector and axial-vector mesons is given by 
\begin{eqnarray}\label{eq:Lag3}
 \mathcal{L}_{4d, \text{eff}}^{(3),\text{meson}} &=& \text{tr} \left\{ (\partial^{\mu} \tilde{v}^{(n)\nu} - \partial^{\nu} \tilde{v}^{(n)\mu}) 
 \left( g_{v^n v^\ell v^m} [ \tilde{v}_{\mu}^{(\ell)},\tilde{v}_{\nu}^{(m)}] +  g_{v^n a^\ell a^m} [ \tilde{a}_{\mu}^{(\ell)},\tilde{a}_{\nu}^{(m)}] \right)\right. \nonumber \\
 && + \left. g_{v^\ell a^m a^n} (\partial^{\mu} \tilde{a}^{(n)\nu} - \partial^{\nu} \tilde{a}^{(n)\mu}) \left([\tilde{v}_{\mu}^{(\ell)},\tilde{a}_{\nu}^{(m)}]-[\tilde{v}_{\nu}^{(\ell)},\tilde{a}_{\mu}^{(m)}] \right)\right\}.
\end{eqnarray}
The 3-meson coupling constants are given by
\begin{eqnarray}
g_{v^\ell v^m v^n} &=& \kappa \int d\tilde{z} C(\tilde{z}) \psi_{2\ell-1} (\tilde{z}) \psi_{2m-1}(\tilde{z}) \psi_{2n-1}(\tilde{z}),\label{eq:gvvv}\\
g_{v^\ell a^m a^n} &=& \kappa \int d\tilde{z} C(\tilde{z}) \psi_{2\ell-1} (\tilde{z}) \psi_{2m}(\tilde{z}) \psi_{2n}(\tilde{z}).\label{eq:gvaa}
\end{eqnarray}
The next step is to numerically calculate the wave functions $\psi_n(\tilde{z})$ in order to be able to obtain the couplings $g_{v^n}$, 
$g_{v^\ell v^m v^n}$, $g_{v^\ell a^m a^n}$ and the masses $M_{v^n}^2$ and $M_{a^n}^2$.

\subsection{Numerical study of the wave functions}\label{sec:wf}
To study and numerically solve the field equations for the vector and axial-vector meson modes, we will employ a shooting method (cf.~\cite{Sakai:2004cn, BallonBayona:2009ar}). In order to find the correct asymptotic behavior of the wave functions $\psi_n(\tilde{z})$, we expand them into a Frobenius series $\psi_n(\tilde{z}) = \tilde{z}^{-\alpha} \sum_m c_{n,m} \tilde{z}^{-m}$ for large $\tilde{z}$. Substituting this expansion into (\ref{eq:eom}), and imposing the normalization condition (\ref{eq:norm}), we find that asymptotically $\psi_n(\tilde{z} \rightarrow + \infty) \sim  \tilde{z}^{-1/2}$.\\
Therefore, it will be convenient to define $\tilde{\psi}_n(\tilde{z}) = \tilde{z}^{1/2} \psi_n(\tilde{z})$. Introducing the new variable $\tilde{z}=e^{\eta}$, the equation of motion for $\tilde{\psi}_n$ can be cast into the following form:
\begin{equation}\label{eq:neweom}
\partial_{\eta}^2 \tilde{\psi}_n + A \partial_{\eta} \tilde{\psi}_n + B \tilde{\psi}_n =0, 
\end{equation}
where
\begin{eqnarray}
A(\tilde{z}=e^{\eta}) &=& - \frac{1+ 4 \tilde{z}^{-2}}{2(1+ \tilde{z}^{-2})}= \sum_{k=0}^{\infty} A_k \tilde{z}^{-k/2},\\
B(\tilde{z}=e^{\eta}) &=& \frac{3}{4} \frac{\tilde{z}^{-2}}{(1+ \tilde{z}^{-2})} + \lambda_n \frac{\tilde{z}^{-1/2}}{(1+ \tilde{z}^{-2})^{5/4}}= \sum_{k=0}^{\infty} B_k \tilde{z}^{-k/2}.
\end{eqnarray}
The first few coefficients are 
\begin{eqnarray}
A_0 = -\frac{1}{2}, A_1=A_2=A_3=0, A_4 = -\frac{3}{2},A_5=A_6=A_7=0, A_8=\frac{3}{2},\ldots \\
B_0 = 0, B_1= \lambda_n, B_2=B_3=0,B_4=\frac{3}{4}, B_5=-\frac{5}{4}\lambda_n,B_6=B_7=0,B_8=-\frac{3}{4},\ldots.
\end{eqnarray}
Finally, by expanding 
\begin{equation}\label{eq:psiexp}
\tilde{\psi}_n (\tilde{z})= \sum_{m=0}^{\infty} \alpha_m \tilde{z}^{-m/2},
\end{equation}
with $\alpha_0=1$, and substituting this expansion into (\ref{eq:neweom}), we find the 
following recursion relation for $\alpha_m$:
\begin{eqnarray}
\alpha_1 &=&-2 B_1,\\ 
\alpha_m &=& \left( \frac{m^2}{4}+\frac{m}{4}\right)^{-1} \left( \frac{1}{2} \sum_{k=1}^{m-1} k A_{m-k} \alpha_k - \sum_{k=0}^{m-1} B_{m-k} \alpha_k\right).
\end{eqnarray}
These relations can now be taken as input data to solve the field equations (\ref{eq:eom}) numerically via a shooting method. 
\subsection{Results}
\begin{figure}[ht] 
\begin{center}
\epsfig{file=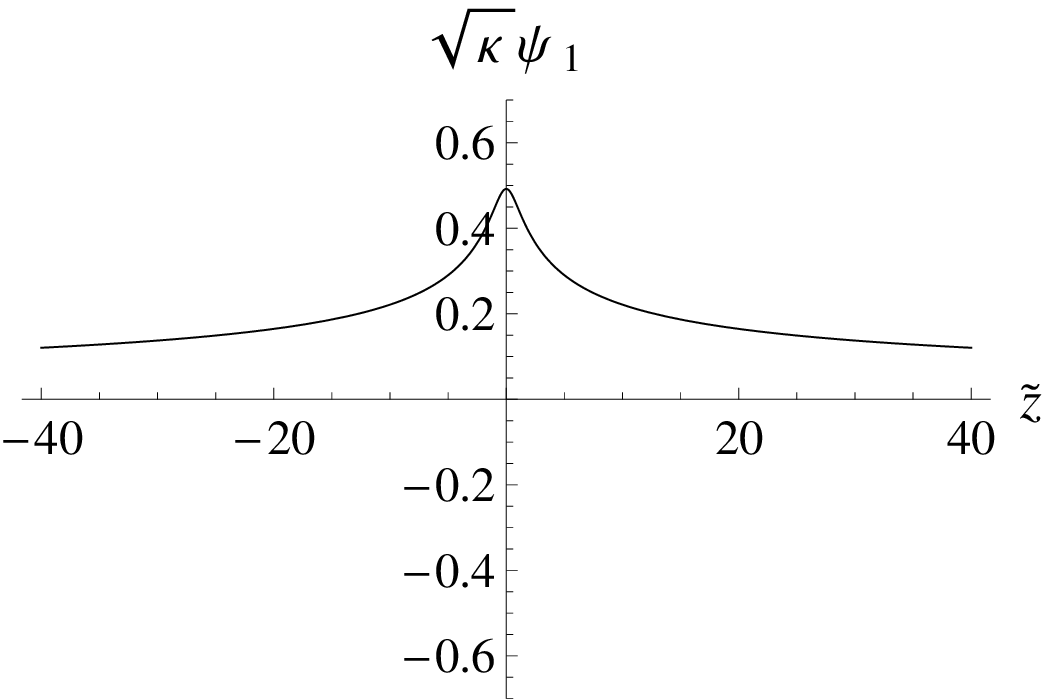, width=6cm}
\epsfig{file=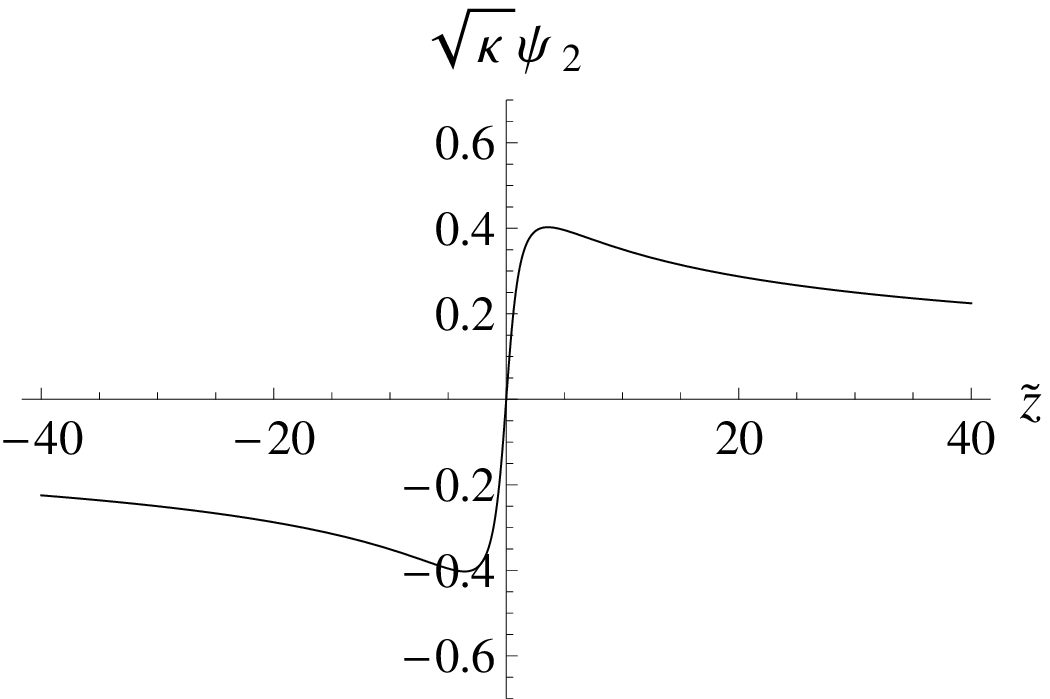, width=6cm}
\epsfig{file=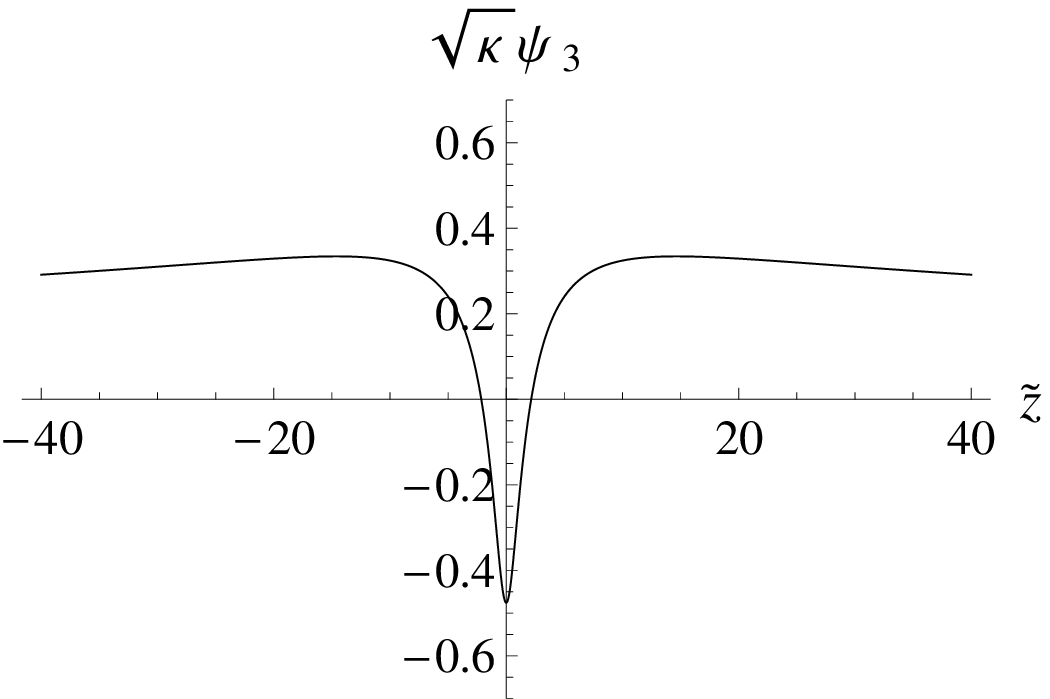, width=6cm}
\epsfig{file=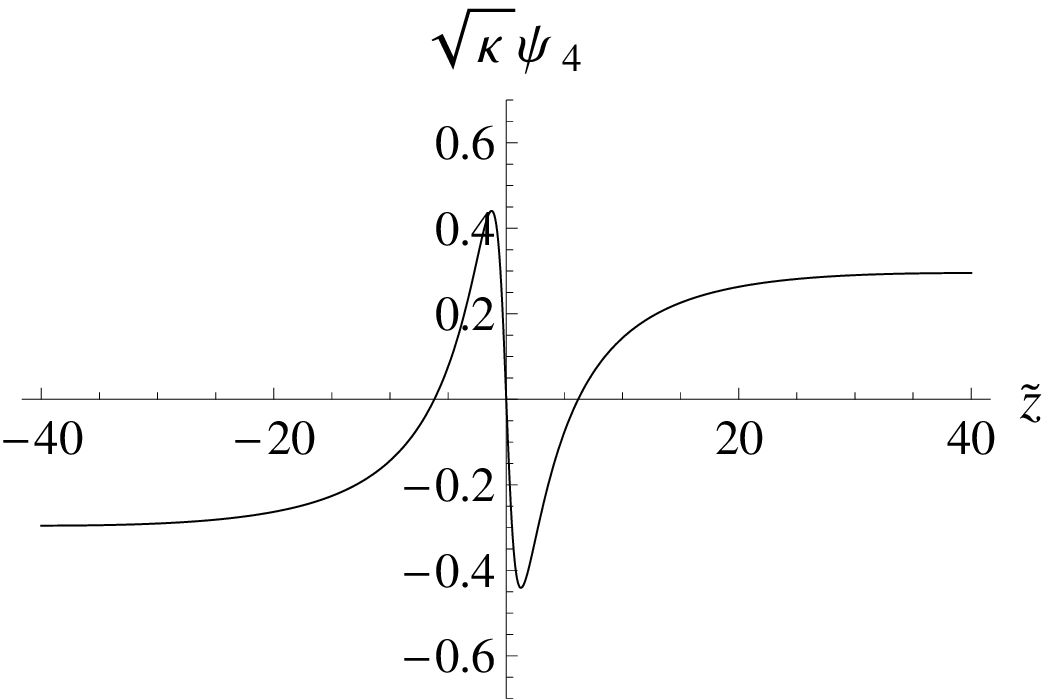, width=6cm}
\epsfig{file=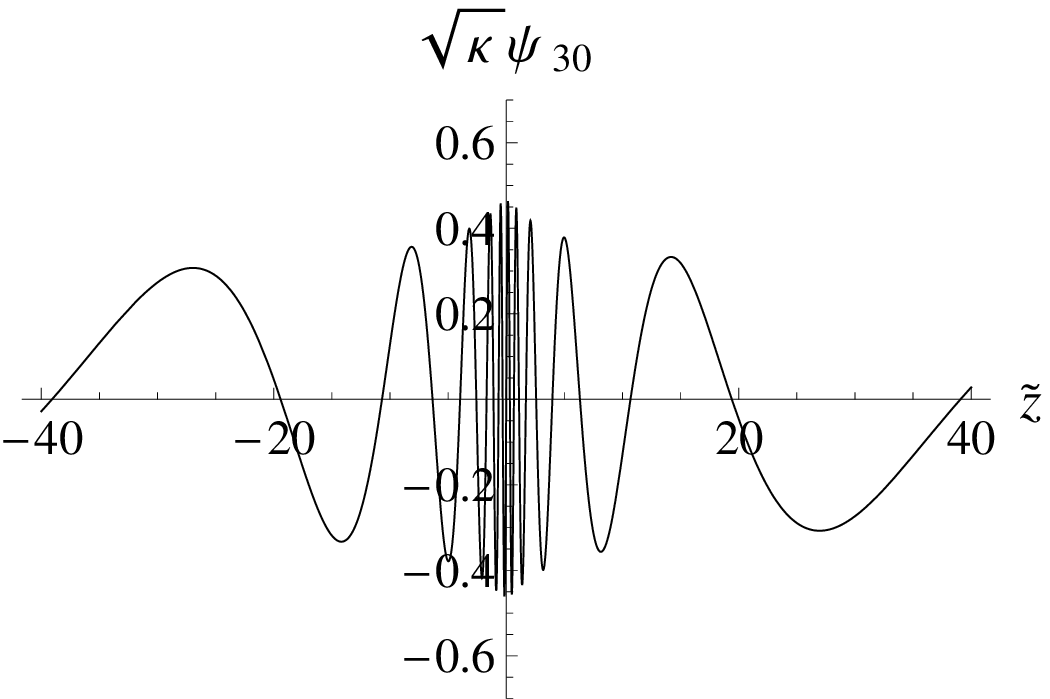, width=6cm}
\epsfig{file=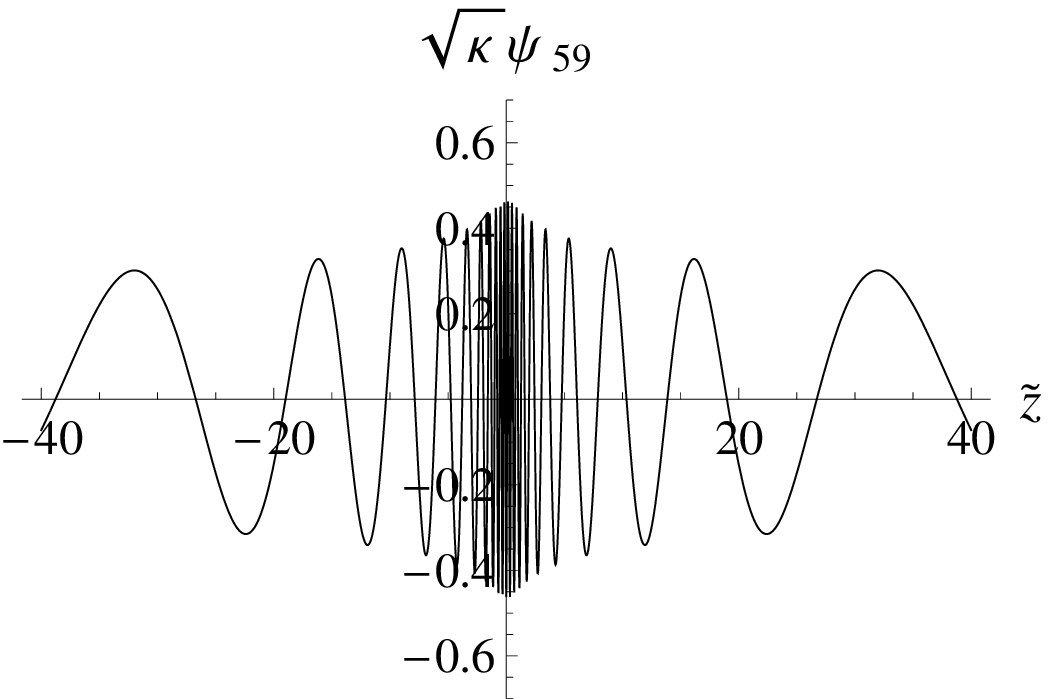, width=6cm}
\end{center}
\caption{Wave functions $\sqrt{\kappa} \psi_n(\tilde{z})$ for $n=1,2,3,4,30,59$.}
\label{fig:wavefct}
\end{figure}

\noindent
{\bf Parity.}
Let us briefly discuss the behavior of the wave functions under the parity transformation $(t,\vec{x},\tilde{z}) \rightarrow (t,-\vec{x},-\tilde{z})$.
A parity transformation exchanges the chiral components of the external gauge fields, $A_L \leftrightarrow A_R$. The vector and axial-vector meson modes
exhibit the following behavior under parity:
\begin{equation}
v_{\mu}^{(n)}(t,-\vec{x}) = + v_{\mu}^{(n)}(t,-\vec{x}), \qquad a_{\mu}^{(n)}(t,-\vec{x}) = - a_{\mu}^{(n)}(t,-\vec{x}).
\end{equation}
To ensure gauge invariance of the five-dimenional gauge fields, we impose the conditions
\begin{equation}
\psi_{2n}(-\tilde{z}) = + \psi_{2n}(\tilde{z}), \qquad  \psi_{2n-1}(-\tilde{z}) = - \psi_{2n-1}(\tilde{z}).
\end{equation}
If we demand regularity of the wave functions $\psi_n(\tilde{z})$ at the origin $\tilde{z}=0$, from the above conditions we conclude that
\begin{equation}
\partial_{\tilde{z}} \psi_{2n}(0) =0, \qquad \psi_{2n-1}(0)=0. 
\end{equation}
Using the large $\tilde{z}$ behavior, eq.~(\ref{eq:psiexp}), and the conditions at $\tilde{z}=0$, we are now able to numerically solve the equations of motion (\ref{eq:eom}). We found solutions $\psi_n(\tilde{z})$ and the corresponding eigenvalues $\lambda_n$ for $n = 1, \ldots, 60$.
The results for a selection of wave functions are shown in figure~\ref{fig:wavefct}. 
For large $\tilde{z}$ one finds the expected $\tilde{z}^{-1/2}$ behavior. In the small $\tilde{z}$ limit, the equations of motion reduce to
\begin{equation}
\psi_n''(\tilde{z})+ \lambda_n \psi_n(\tilde{z})=0,
\end{equation}
which has sinusoidal solutions
\begin{equation}
\psi_{2n} \sim \sin (\sqrt{\lambda_{2n}} \tilde{z}), \qquad \psi_{2n-1} \sim \cos (\sqrt{\lambda_{2n-1}} \tilde{z}) 
\end{equation}
showing oscillatory behavior. This is compatible with the observed behavior of $\psi_n(\tilde{z})$ in the small $\tilde{z}$ regime.
The corresponding mass eigenvalues are summarized in table~\ref{tab:lambda}. The numerical values found here are consistent with the values quoted in \cite{Kuperstein:2008cq}, eq.~(4.24).\footnote{In order to compare the numerical values, it is important to observe that the $\lambda_n$ of \cite{Kuperstein:2008cq} in reality are 
differently normalized meson masses $M_n$. For direct comparison between the two papers, one has to use the relation $\lambda_{n,\text{here}}= \frac{\lambda_{n,\text{KS}}^2}{16}$.}

\begin{table}[h]
\begin{center}
\begin{tabular}{|c||c|c|c|c|c|c|c|c|c||}
\hline
$n$ & 1 & 2 & 3 & 4 & 5 & 6 & 7 & 8 & 9 \\\hline
$\lambda_{2n-1}=\frac{M_{v^n}^2}{M_{\ast}^2}$ & 0.258 & 1.384 & 3.432 & 6.393 & 10.266 & 15.049 & 20.745 & 27.348 & 34.866 \\\hline
$\lambda_{2n}=\frac{M_{a^n}^2}{M_{\ast}^2}$ & 0.691 & 2.292 & 4.798 & 8.215 & 12.408 & 17.783 & 23.932 & 30.995 & 38.920 \\\hline
\end{tabular}
\end{center}
\caption{Some numerical values for the dimensionless vector and axial-vector masses.}\label{tab:lambda}
\end{table}
\noindent
The lightest meson, $v^{(1)}$, is a vector meson that could, for the sake of comparison with QCD, be identified with the $\rho$-meson $\rho(770)$. The second lightest meson, $a^{(1)}$, is an axial-vector meson identifiable with $a_1(1260)$, the third lightest, $v^{(2)}$, a vector meson identifiable with $\rho(1450)$ and so on.
In table \ref{tab:ratio}, we compare the mass ratios of the first few mesons in the KS model to the Sakai-Sugimoto (SS) model and experiments (QCD). It is important to keep in mind, however, that the predictions of any gauge-gravity model are stricly speaking only valid in the large $N_c$, large t'Hooft coupling limit and that we are not studying a QCD-dual here, but rather a dual to the quiver gauge theory described above.

\begin{table}[h]
\begin{center}
\begin{tabular}{|c||c|c|c|}
\hline
k & $\left(\frac{\lambda_{k+1}}{\lambda_1}\right)_{\text{KS}}$  & $\left(\frac{\lambda_{k+1}}{\lambda_1}\right)_{\text{SS}}$ & $\left(\frac{\lambda_{k+1}}{\lambda_1}\right)_{\text{QCD}}$ \\\hline
1 & 2.68 & 2.4 & $\sim$ 2.51 \\\hline
2 & 5.36 & 4.3 & $\sim$ 3.56 \\\hline
\end{tabular}
\end{center}
\caption{Ratios $M_{k+1}^2/M_1^2 = \lambda_{k+1}/\lambda_1$ for $k=1,2$. $M_1^2$ corresponds to the mass squared of the $\rho$-meson, which can be matched in our model by setting $M_{\ast}^2= (1527 \text{MeV})^2$. This can be done by adjusting $r_0^2$ appropriately.}\label{tab:ratio}
\end{table}

\begin{figure}[ht]
 \epsfig{file=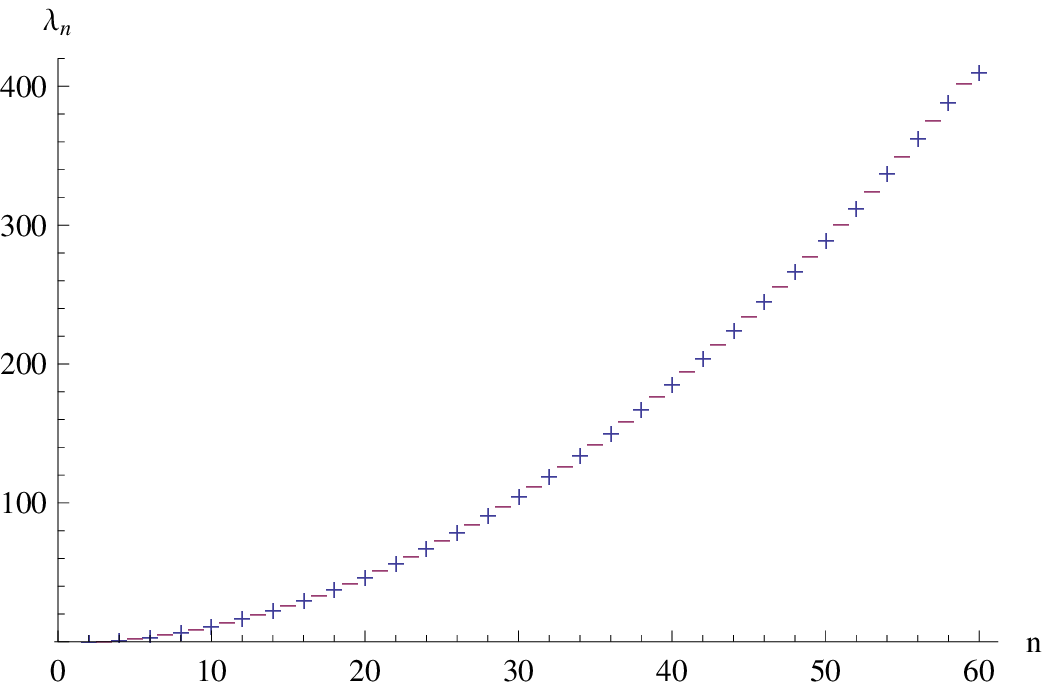, width=7.5cm}
 \epsfig{file=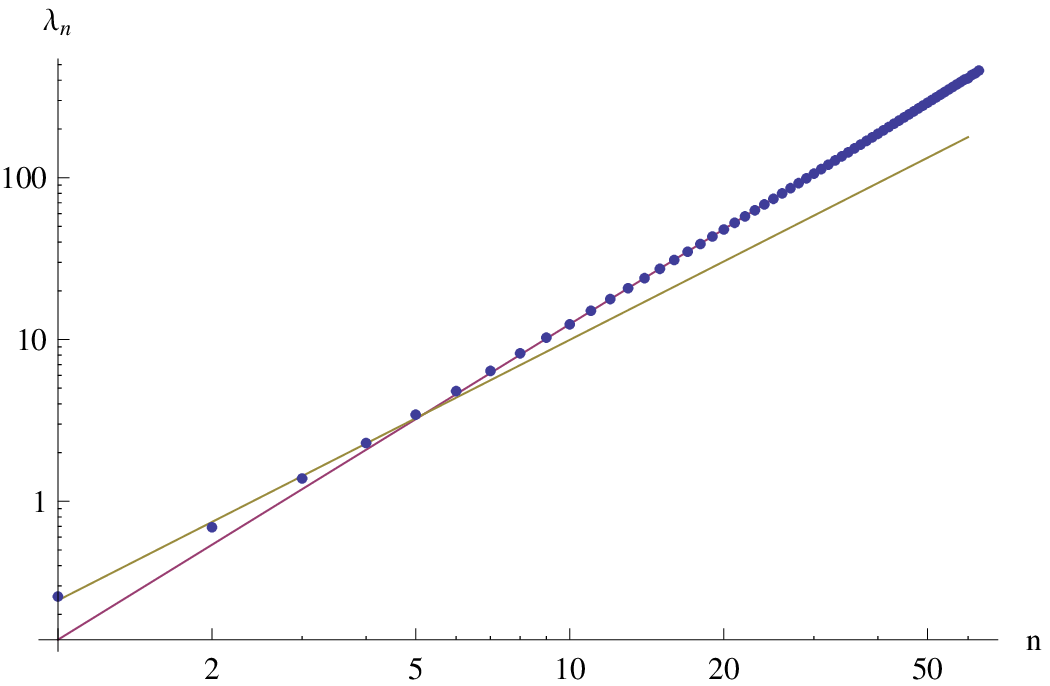, width=7.5cm}
\caption{Regge trajectory for vector and axial vector mesons. In the left graph, $\lambda_n$ is plotted versus $n$, for both vector mesons (+) and axial vector mesons (-). The right graph is a logarithmic plot of both vector and axial-vector mesons. The straight lines are two linear best fits for small $n\leq 5$ and $n=6, \ldots 60$.}
\label{fig:Regge}
\end{figure}

\noindent
{\bf Regge trajectories.}
Next, we examine the dependence of the eigenvalues $\lambda_n$ on the radial number $n$, which we will refer to as a Regge trajectory. These eigenvalues determine the meson masses according to $M_n^2 = \lambda_n M_{\ast}^2$. The Regge trajectory for both the vector and axial-vector mesons is shown in figure~\ref{fig:Regge}. We also display a logarithmic plot with linear best fits for both small $n=1,\ldots,5$ and large $n=6,\ldots,60$.
In the small $n$ region, the best fit is $-1.41+1.61 \log n$, corresponding to $\lambda_n \sim n^{1.61}$, while in the large $n$ region, we find 
$-1.97+1.95 \log n$, corresponding to $\lambda_n \sim n^{1.95}$. This result is consistent with previous results presented in the literature, where the Regge trajectory for vector and axial-vector mesons was found to exhibit asymptotic quadratic behavior $\lambda_n \sim n^2$ in the $D4-D8$ model \cite{BallonBayona:2009ar}, the hard-wall model \cite{BoschiFilho:2002ta,BoschiFilho:2002vd,deTeramond:2005su,Erlich:2005qh,BoschiFilho:2005yh} and the $D3-D7$ model \cite{Kruczenski:2003be,Kirsch:2006he}. Thus, this seems to be a general feature of holographic models of mesons.\\\\
\noindent
{\bf Decay and coupling constants.}
We are now ready to calculate the decay and coupling constants. Since we will be interested in studying (elastic and non-elastic) form factors, i.e. interactions of particles with external photons, which naturally couple to vector mesons, we will focus here on the vector meson decay constant $g_{v^n}$ and the coupling constants between the intermediate vector meson and the external vector mesons $g_{v^n v^l v^m}$, and between the intermediate 
vector meson and the external axial-vector mesons $g_{v^n a^l a^m}$, respectively. 
\begin{table}[h]
\begin{center}
\begin{tabular}{|c||c|c|c|c|c|c|c|c|c||}
\hline
$n$ & 1 & 2 & 3 & 4 & 5 & 6 & 7 & 8 & 9 \\\hline\hline
$\frac{g_{v^n}}{\sqrt{\kappa}M_{\ast}^2}$ & \tiny{0.829} & \tiny{2.965} & \tiny{5.864} & \tiny{9.347} & \tiny{13.331} & \tiny{17.759} & \tiny{22.591} & \tiny{27.793}& \tiny{33.345}  \\\hline\hline
$\sqrt{\kappa}g_{v^nv^1v^1}$  & \tiny{0.383} & \tiny{-0.110} & \tiny{0.00176} & \tiny{-0.000210} & \tiny{0.0000631} & \tiny{-0.0000194} & \tiny{0.0000265} & \tiny{0.0000311} & \tiny{-0.0000213} \\\hline
$\sqrt{\kappa}g_{v^nv^1v^2}$  & \tiny{-0.110} & \tiny{0.258} & \tiny{-0.119} & \tiny{0.00254} & \tiny{-0.000349} & \tiny{0.0000621} & \tiny{-0.0000324} & \tiny{-0.0000320} & \tiny{0.0000225} \\\hline
$\sqrt{\kappa}g_{v^nv^2v^2}$  & \tiny{0.258} & \tiny{0.0579} & \tiny{0.136} & \tiny{-0.131} & \tiny{0.00382} & \tiny{-0.000526} & \tiny{0.000108} & \tiny{0.0000210} & \tiny{-0.0000217} \\\hline\hline
$\frac{g_{a^n}}{\sqrt{\kappa}M_{\ast}^2}$ & \tiny{1.783} & \tiny{4.337} & \tiny{7.539} & \tiny{11.281} & \tiny{15.294} & \tiny{20.127} & \tiny{25.147} & \tiny{30.529} & \tiny{36.243} \\\hline\hline
$\sqrt{\kappa}g_{v^na^1a^1}$  & \tiny{0.270} & \tiny{0.156} & \tiny{-0.122} & \tiny{0.00351} & \tiny{-0.0000443} & \tiny{0.0000673} & \tiny{-0.0000103} & 
\tiny{-0.00000183} & \tiny{0.00000236}  \\\hline
$\sqrt{\kappa}g_{v^na^1a^2}$  & \tiny{-0.115} & \tiny{0.147} & \tiny{0.138} & \tiny{-0.128} & \tiny{0.00384} & \tiny{-0.000507} & \tiny{0.0000805} & \tiny{-0.00000579} & \tiny{-0.00000342}   \\\hline
$\sqrt{\kappa}g_{v^na^2a^2}$  & \tiny{0.249} & \tiny{0.0540} & \tiny{0.0484} & \tiny{0.117} & \tiny{-0.136} & \tiny{0.00422} & \tiny{-0.000594} & \tiny{0.0000829} & \tiny{0.00000500}  \\\hline\hline
\end{tabular}
\end{center}
\caption{Dimensionless decay and coupling constants for vector and axial-vector mesons.}\label{tab:vaconst}
\end{table}
 The calculation was carried out using eqs.~(\ref{eq:gv}), (\ref{eq:gvvv}) and (\ref{eq:gvaa}), as well as the numerical results for the normalized wave functions found above, for $n=1, \ldots, 60$. The results of the computations are shown in table \ref{tab:vaconst} for both vector and axial-vector mesons.
For completeness, we also include the coupling constants $g_{a^n}$. 
\section{Pions}\label{sec:pion}
\noindent
{\bf Skyrme model.}
Let us briefly discuss the effective Lagrangian for the pions coming from the DBI part of the $D7$-brane action. To this end, it is convenient to omit the meson fields $v_{\mu}^{(n)}$, $a_{\mu}^{(n)}$, 
and to work in a gauge where $A_{\tilde{z}}=0$, $\xi_-(x^{\mu}) =1$ and $\xi_+^{-1} (x^{\mu}) = U(x^{\mu}) := e^{i \frac{\Pi(x)}{f_\pi}}$. Then we are left with the mode expansion 
\begin{equation}
A_{\mu}(x^{\mu},\tilde{z}) = U^{-1}(x^{\mu}) \partial_{\mu} U(x^{\mu}) \psi_+ (\tilde{z}),
\end{equation}
which leads to
\begin{eqnarray}
F_{\mu \nu} &=& \left[U^{-1}\partial_{\mu} U, U^{-1} \partial_{\nu} U \right] \psi_+ (\psi_+ -1),\\
F_{\tilde{z} \mu} &=& U^{-1}\partial_{\mu} U \partial_{\tilde{z}} \psi_+.
\end{eqnarray}
The resulting action for the pions can then be written as
\begin{equation}
S_{5d,\text{eff}}^{\text{pion}} = \kappa \int d^4 x \int d\tilde{z} \, \text{tr} \left( \frac{1}{2} C(\tilde{z}) \psi_+^2 (\psi_+ -1)^2  \left[U^{-1}\partial_{\mu} U, U^{-1} \partial_{\nu} U \right]^2 + M_{\ast}^2 D(\tilde{z}) (U^{-1}\partial_{\mu} U)^2 (\partial_{\tilde{z}} \psi_+)^2 \right). 
\end{equation}
This is precisely the action of the Skyrme model
\begin{equation}
 S_{\text{Skyrme}}= \int d^4 x \left( \frac{f_{\pi}^2}{4} \text{tr} (U^{-1}\partial_{\mu} U)^2 + \frac{1}{32 e_S^2} \text{tr} \left[U^{-1}\partial_{\mu} U, U^{-1} \partial_{\nu} U \right]^2\right),
\end{equation}
with the following identifications: The pion decay constant $f_{\pi}$ ($\sim$ 93 MeV experimentally in QCD) is given by
\begin{equation}\label{eq:fpi}
 f_{\pi}^2 = 4 \kappa M_{\ast}^2 \int d\tilde{z} \, D(\tilde{z})  (\partial_{\tilde{z}} \psi_+)^2 =\frac{4 \Gamma (3/4)}{\sqrt{\pi}\Gamma (1/4)}\kappa M_{\ast}^2 \approx 0.76 \kappa M_{\ast}^2,
\end{equation}
and the dimensionless parameter $e_S$ is calculated as
\begin{equation}
 e_S^{-2} = 32 \kappa \int d\tilde{z} \frac{1}{2} C(\tilde{z}) \psi_+^2 (\psi_+ -1)^2 = \kappa \int d\tilde{z} C(\tilde{z}) (1-\psi_0^2)^2 \approx 3.93 \kappa.
\end{equation}
Here we used the definition of $\psi_+ (\tilde{z})= \frac{1}{2} (1 + \psi_0 (\tilde{z}))$. In order to match the experimental value for $f_{\pi}$, the 
dimensionless constant $\kappa$ has to be adjusted accordingly:
\begin{equation}
\kappa \approx 4.87 \times 10^{-3}. 
\end{equation}
In large $N_c$ QCD, one expects $f_{\pi}^2 \sim \mathcal{O} (N_c)$ and $e_S^{-2} \sim \mathcal{O} (N_c)$, which is consistent with the dependence on
$\kappa$ (cf. the expression below eq.~(\ref{eq:5daction})).\\

\noindent
{\bf Interaction terms involving pions.} So far, we have only studied the part of the interaction Lagrangian $ \mathcal{L}_{4d, \text{eff}}^{(3)}$ 
involving mesons. For the computation of the pion form factor below, we also need to introduce the part involving pions and mesons simultaneously. To this 
end, we rewrite the kinetic term for the vector mesons using the redefinition
\begin{equation}
\widehat{v}_{\mu}^{(n)}=  \tilde{v}_{\mu}^{(n)} + \frac{g_{v^n \pi \pi}}{M_{v^n}^2} [\Pi,\partial_{\mu} \Pi].
\end{equation}
Now, written in a form closely resembling the effective Lagrangian given in the literature (cf.~\cite{Sakai:2005yt,Skyrme:1961vq}), the relevant terms are 
\begin{eqnarray*}
 \mathcal{L}_{4d, \text{eff}}^{(2)}+ \mathcal{L}_{4d, \text{eff}}^{(3),\text{meson}+\Pi}&=& \frac{1}{2} \text{tr} (\partial_{\mu} \widehat{v}_{\nu}^{(n)} - \partial_{\nu} \widehat{v}_{\mu}^{(n)})^2 
 + \frac{1}{2} \text{tr} (\partial_{\mu} \tilde{a}_{\nu}^{(n)} - \partial_{\nu} \tilde{a}_{\mu}^{(n)})^2 + \text{tr} (i \partial_{\mu} \Pi + f_{\pi} \mathcal{A}_{\mu})^2 \\\nonumber
&& + M^2_{v^n}  \text{tr} \left( \widehat{v}_{\mu}^{(n)} - \frac{g_{v^n}}{M^2_{v^n}} \mathcal{V}_{\mu}\right)^2 + M^2_{a^n}  \text{tr} \left( \tilde{a}_{\mu}^{(n)} - \frac{g_{a^n}}{M^2_{a^n}} \mathcal{A}_{\mu}\right)^2 \\\nonumber
&& - 2 g_{v^n \pi \pi} \text{tr} (\widehat{v}_{\mu}^{(n)}[\Pi,\partial^{\mu} \Pi]) + \left(2 \frac{g_{v^n} g_{v^n \pi \pi}}{M_{v^n}^2}-2\right)
\text{tr} (\mathcal{V}_{\mu} [\Pi,\partial^{\mu}\Pi]) \\\nonumber
&& + \frac{g_{v^n \pi \pi}^2}{M_{v^n}^2} \text{tr} [\Pi,\partial_{\mu} \Pi]^2 - 2 \frac{g_{v^n \pi \pi}^2}{M_{v^n}^4} \text{tr} [\partial_{\mu} \Pi,\partial_{\nu} \Pi]^2.
\end{eqnarray*}
Here we have introduced the coupling constants
\begin{eqnarray}\label{eq:gvpp}
 g_{v^n \pi \pi} &=& \kappa \frac{M_{v^n}^2}{2 f_{\pi}^2}\int_{-\infty}^{+\infty} d\tilde{z}\, C(\tilde{z}) \psi_{2n-1}(\tilde{z}) (1-\psi_0^2),\nonumber \\
 &=& - \int_0^{+\infty} d\tilde{z} \, \psi_0(\tilde{z}) \partial_{\tilde{z}} \psi_{2n-1} (\tilde{z}).
\end{eqnarray}
\noindent
{\bf Vector meson dominance.} 
Using the equation of motion (\ref{eq:eom}), together with the completeness relation 
\begin{equation}
\kappa \sum_{n=1}^{\infty} C(\tilde{z}') \psi_{n}(\tilde{z}) \psi_{n}(\tilde{z}') = \delta(\tilde{z}-\tilde{z}'), \label{eq:completeness}
\end{equation}
one can prove the important sum rule\footnote{This sum rule turns out to be universal in holographic models. It has be shown to hold in the soft-wall and hard-wall models \cite{Grigoryan:2007vg}, $D3/D7$-models \cite{Hong:2004sa} and the Sakai-Sugimoto $D4/D8$-model \cite{Sakai:2005yt}.}
\begin{equation}\label{eq:pionvmd}
 \sum_{n=1}^{\infty} \frac{g_{v^n}}{M_{v^n}^2} g_{v^n \pi \pi} = 1.
\end{equation}
This shows that the $\mathcal{V}\pi\pi$ coupling in the above Lagrangian vanishes, leading to vector meson dominance in the pion form factor. This result descends from a more general class of geometries and brane embeddings leading to a five dimensional effective action as in (\ref{eq:5daction}). General $C(\tilde{z})$ and $D(\tilde{z})$ terms need to obey a few conditions in order to reproduce vector meson dominance and other properties of pion and meson phenomenology. A finite pion decay constant $f_{\pi}$ requires that (cf.~(\ref{eq:fpi}))
\begin{equation}
 f_\pi \sim \int d\tilde{z} \frac{1}{D(\tilde{z})} 
\end{equation}
be finite. Renormalizable five dimensional gauge fields require vanishing wave functions at $\tilde{z}\rightarrow \pm \infty$ whose behavior depends on $C(\tilde{z})$ and $D(\tilde{z})$ in (\ref{eq:norm}). Parity requires that both $C(\tilde{z})$ and $D(\tilde{z})$ be even functions of $\tilde{z}$. Noticing that the equations of motion (\ref{eq:norm}) are of Sturm-Liouville type, the above requirements set the conditions for an orthonormal set of wave functions $\psi_n(\tilde{z})$. Vector meson dominance hence follows from applying the wave functions completeness relation (\ref{eq:completeness}) and equation of motion (\ref{eq:eom}) to the sum rule expressions for the pions (\ref{eq:pionvmd}) and mesons (\ref{eq:mesonvmd}).  In \cite{Hong:2004sa}, the authors claim that vector meson dominance is a feature of asymptotic AdS spaces, but it should be noted that it is in fact more general and also applies to other geometries, e.g., the Sakai-Sugimoto model. \\  
Similar to (\ref{eq:pionvmd}), one can prove two more sum rules:
\begin{equation}\label{eq:sumrulespion}
\sum_{n=1}^{\infty} \frac{g_{v^n \pi \pi}^2}{M_{v^n}^2} = \frac{1}{3 f_{\pi}^2}, \quad \sum_{n=1}^{\infty} \frac{g_{v^n \pi \pi}^2}{M_{v^n}^4}
= \frac{1}{4} \frac{1}{e_S^2 f_{\pi}^4}.
\end{equation}
In conclusion, the kinetic and interaction Lagrangian involving both mesons and pions can now be written in its final form:
\begin{eqnarray}\label{eq:pionLag}
 \mathcal{L}_{4d, \text{eff}}^{(2)}+ \mathcal{L}_{4d, \text{eff}}^{(3),\text{meson}+\Pi}&=& \frac{1}{2} \text{tr} (\partial_{\mu} \widehat{v}_{\nu}^{(n)} - \partial_{\nu} \widehat{v}_{\mu}^{(n)})^2 
 + \frac{1}{2} \text{tr} (\partial_{\mu} \tilde{a}_{\nu}^{(n)} - \partial_{\nu} \tilde{a}_{\mu}^{(n)})^2  \nonumber \\
&& + \text{tr} (i \partial_{\mu} \Pi + f_{\pi} \mathcal{A}_{\mu})^2 + M^2_{v^n}  \text{tr} \left( \widehat{v}_{\mu}^{(n)} - \frac{g_{v^n}}{M^2_{v^n}} \mathcal{V}_{\mu}\right)^2  \nonumber \\
&& + M^2_{a^n}  \text{tr} \left( \tilde{a}_{\mu}^{(n)} - \frac{g_{a^n}}{M^2_{a^n}} \mathcal{A}_{\mu}\right)^2 - 2 g_{v^n \pi \pi} \text{tr} (\widehat{v}_{\mu}^{(n)}[\Pi,\partial^{\mu} \Pi]) \nonumber \\
&& + \frac{1}{3 f_{\pi}^2} \text{tr} [\Pi,\partial_{\mu} \Pi]^2 - \frac{1}{2 e_S^2 f_{\pi}^4} \text{tr} [\partial_{\mu} \Pi,\partial_{\nu} \Pi]^2.
\end{eqnarray}
The wave functions calculated in section \ref{sec:wf} fall off rather slowly at large $\tilde{z}$. This results in fairly slow convergence
of the relevant integrals involved in studying the pion form factor. For the results presented below, we have integrated numerically up to $\tilde{z}_{\text{max}}=10^{21}$ using Mathematica. With this precision, we can roughly trust the first 15 coupling constants 
$g_{v^n \pi \pi}$, $n=1, \ldots, 15$ introduced in eq.~(\ref{eq:gvpp}).
In table \ref{tab:gvpp}, we list some numerical results for these coupling constants. 
\begin{table}[h]
\begin{center}
\begin{tabular}{|c||c|c|c|c|c|c|c|c|c|c||}
\hline
$n$ & 1 & 2 & 3 & 4 & 5 & 6 & 7 & 8 & 9 \\\hline
$\sqrt{\kappa}g_{v^n \pi \pi}$  & \tiny{0.335} & \tiny{-0.0425} & \tiny{0.00914} & \tiny{-0.00216} & \tiny{0.000545} & \tiny{-0.000141} & \tiny{0.0000535} & \tiny{0.0000155} & \tiny{-0.0000140} \\\hline\hline
\end{tabular}
\end{center}
\caption{Coupling constants for a vector meson $v^n$ coupling to pions.}\label{tab:gvpp}
\end{table}

\section{Form factors}\label{sec:form}

\subsection{Pion form factor}
The electromagnetic form factor represents the interaction of a particle with an external photon.
In this section, we will study the pion form factor $F_{\pi}(q^2)$ (cf. figure 3), 
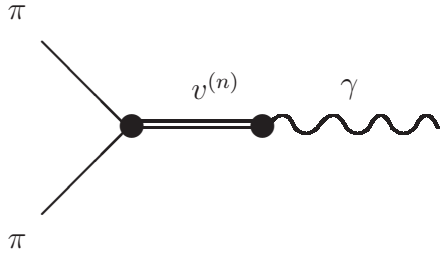
\begin{figure}[h]%\label{fig:pionff}
\begin{center}
\vskip 3.cm
\begin{picture}(0,0)(5,0)
\setlength{\unitlength}{0.06in}
\rm
\thicklines 
%%%%%%%%%%%%%%%%%%%%%%%%%%%%%%%%%%%%%%%%%  Pions %%%%%%%%%%%%%%%%%%%%%%
\put(-12,17){$\pi$}
%\multiput(-9,15)(4,0){13}{\line(1,0){2}}
\put(-9,15){\line(1,-1){7.5}}
\put(-12,-3){$\pi$}
\put(-9,0){\line(1,1){7.5}}
\put(10.2,7.7){\circle*{2}}
\put(-1.2,8.1){\line(1,0){10.5}}
\put(-1.2,7.5){\line(1,0){10.5}}
\put(-1.2,7.7){\circle*{2}}
\put(4,10){$v^{(n)}$}
%%%%%%%%%%%%%%%%%%%%%%%%%%%%%%%%%%%%%%%%%%%%%%%%%  Photon  %%%%%%%%%%%%%%%%%%%%%
\put(17,10.5){$\gamma$}
\bezier{300}(10.5,7.5)(12.2,9.7)(13,7.5)
\bezier{300}(13,7.5)(14,6.5)(15,7.5)
\bezier{300}(15,7.5)(16.5,9.7)(17.5,7.5)
\bezier{300}(17.5,7.5)(18.5,6.5)(19.5,7.5)
\bezier{300}(19.5,7.5)(20.5,9.7)(21.5,7.5)
\bezier{300}(21.5,7.5)(22.5,6.5)(23.5,7.5)
\bezier{300}(23.5,7.5)(24.5,9.7)(25.5,7.5)
\end{picture}
\vskip 1.cm
\parbox{4.1 in}{\caption{Feynman diagram for the pion form factor, showing vector meson dominance.}}
\end{center}
\end{figure}
which is defined by the following matrix element of the electromagnetic current,
\begin{equation}
 \langle \pi^a(p) \vert J^{\mu c}(0) \vert \pi^b(p') \rangle = f^{abc} (p+p')^{\mu} F_{\pi}((p-p')^2), 
\end{equation}
where $J^{\mu}$ is the conserved vector current coupled to $\mathcal{V}_{\mu}$ and $f^{abc}$ are the structure constants of $U(N_f)$. 
By combining the relevant vertices and propagators (cf. eq.~(\ref{eq:pionLag})), according to figure 3, we arrive at the following expression for the pion form factor,
\begin{equation}\label{eq:pionff}
F_{\pi} (q^2) = \sum_{n=1}^{\infty} \frac{g_{v^n} g_{v^n \pi \pi}}{q^2 + M_{v^n}^2}. 
\end{equation}
Comparing eq.~(\ref{eq:pionvmd}) to eq.~(\ref{eq:pionff}), one finds that $F_{\pi}(0) =1$. We can use this as a test of the quality of our numerics.
Summing up the first 15 terms in (\ref{eq:pionff}), we find
\begin{equation}
F_{\pi} (0) \approx \sum_{n=1}^{15} \frac{g_{v^n}}{M_{v^n}^2} g_{v^n \pi \pi} \approx 0.999902.
\end{equation}
The numerical results for the pion form factor are presented in figure \ref{fig:pffgraphs}. 
\begin{figure}[ht]
 \epsfig{file=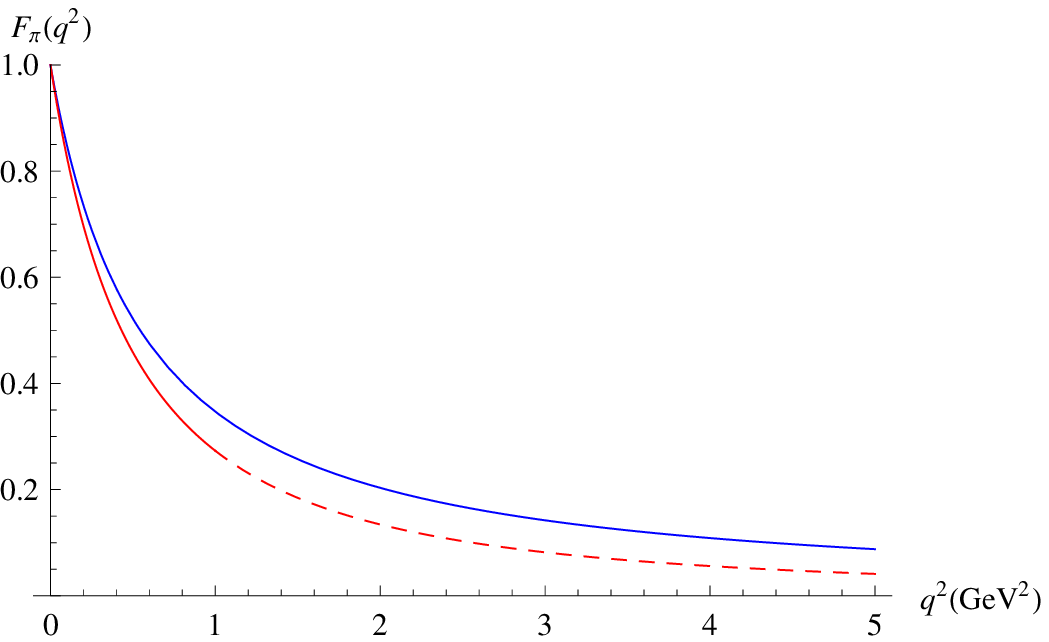, width=8cm}
 \epsfig{file=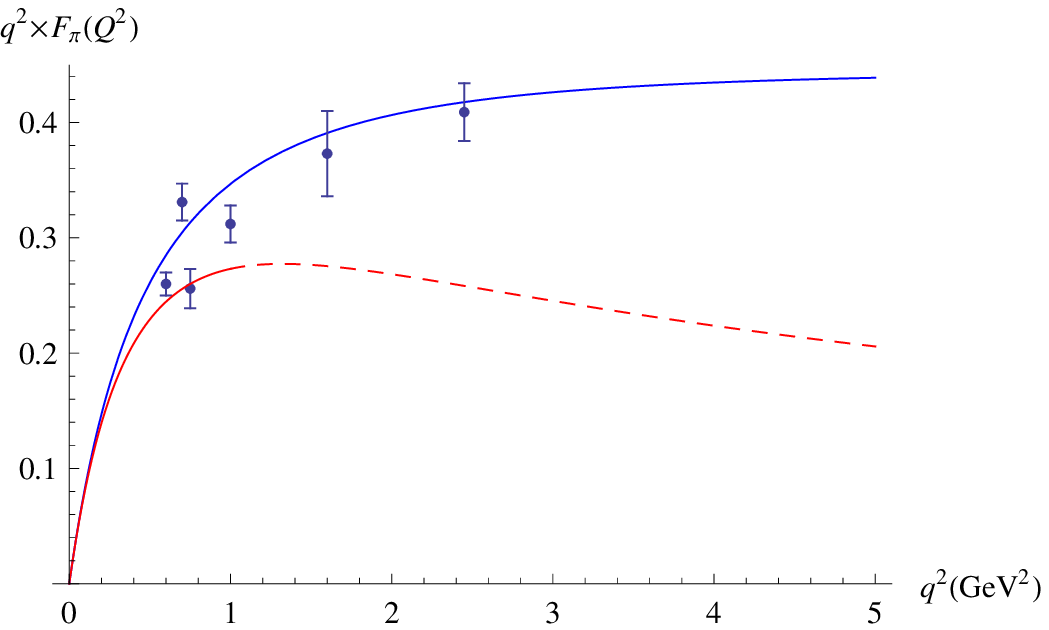, width=8cm}
\caption{The plot on the left shows the pion form factor for the KS model (solid blue line). On the right, we present a plot of $q^2 \times F_{\pi}(q^2)$, 
which exhibits the expected $q^{-2}$ behavior at large $q^2$. The KS model (solid blue line) agrees very well with the data obtained from QCD experiments, taken from \cite{Horn:2006tm,Tadevosyan:2007yd}. The error bars represent all experimental and analysis errors, but not ``model uncertainty'', as explained in the aforementioned papers. For comparison, we also included the results for the Sakai-Sugimoto model (solid red lines), compiled using data from \cite{Sakai:2005yt,BallonBayona:2009ar}. Note that the results for the Sakai-Sugimoto model should only be trusted up to the KK scale $M_{KK}^2 \sim (0.946 \text{GeV})^2$. The dashed red lines represent extrapolations based on our numerical calculations in the Sakai-Sugimoto model.}
\label{fig:pffgraphs}
\end{figure}

\noindent
To study the large $q^2$ behavior of the form factor, we perform an expansion in powers of $q^{-2}$:
\begin{equation}
 F_{\pi} (q^2) = \frac{1}{q^2} \sum_{n=1}^{\infty} g_{v^n} g_{v^n \pi \pi} \left( 1 - \frac{M_{v^n}^2}{q^2} + \mathcal{O}(q^{-4}) \right).
\end{equation}
We can calculate the coefficients of the leading $q^{-2}$ term, using our numerical results:
\begin{equation}
\sum_{n=1}^{15} g_{v^n} g_{v^n \pi \pi} \approx 0.443 (\text{GeV})^2,
\end{equation}
which is (to a high accuracy) the asymptotic value of $q^2 \times F_{\pi}(q^2)$.
The large $q^2$ behavior of $F_{\pi}(q^2)\sim q^{-2}$ is a reflection of the fact that pions (as mesons) are made up of two constituents (see the discussion in \cite{Brodsky:2007hb}, appendix D, specifically eq.~(D11)). Note that the large $q^2$ behavior corresponds to $M_{\ast}^2 \rightarrow 0$, i.e., taking the conformal limit $r_0\rightarrow 0$ in the KS model. It is also interesting to observe that for large $q^2$, the slope of $q^2 \times F_{\pi}(q^2)$ in figure \ref{fig:pffgraphs} depends on $\sum_n g_{v^n} g_{v^n \pi \pi} M_{v^n}^2$, which converges to zero quickly in the KS case, signaling conformal behavior, while in the Sakai-Sugimoto model it turns out to be nonzero and negative ($\approx -2.5 (\text{GeV})^4$). Its slope will still eventually converge to zero for very large $q^2$ since it comes with a factor of $q^{-4}$.\\
We can also numerically check the validity of the sum rules (\ref{eq:sumrulespion}),
\begin{equation}
f_{\pi}^2 \sum_{n=1}^{15} \frac{g_{v^n \pi \pi}^2}{M_{v^n}^2} \approx 0.333285, \quad e_S^2 f_{\pi}^4\sum_{n=1}^{15} \frac{g_{v^n \pi \pi}^2}{M_{v^n}^4}
\approx 0.249927.
\end{equation}
Moreover, we are now in a position to estimate the charge radius of the pion. Expanding $F_{\pi}(q^2)$ for small $q^2$ as 
\begin{equation}
F_{\pi}(q^2) = 1- \sum_{n=1}^{\infty}\frac{g_{v^n} g_{v^n \pi \pi}}{M_{v^n}^4} q^2 + \mathcal{O}(q^4), 
\end{equation}
we compute
\begin{equation}
\langle r^2_{\pi} \rangle = - 6 \frac{d^2}{d q^2} F_{\pi}(q^2)\Big{|}_{q^2=0} = 6 \sum_{n=1}^{\infty}\frac{g_{v^n} g_{v^n \pi \pi}}{M_{v^n}^4}. 
\end{equation}
Summing up the first 15 terms, as before, yields
\begin{equation}
 \langle r^2_{\pi} \rangle \approx 24.707 M_{\ast}^{-2} = (0.641 \text{fm})^2,
\end{equation}
which is close to the experimental value of $\langle r^2_{\pi} \rangle_{\text{exp}} \approx (0.672 \text{fm})^2$.

\subsection{Vector and axial-vector meson form factors}
Here we will calculate the form factors for vector and axial-vector mesons, again following the strategy outlined in \cite{BallonBayona:2009ar} for the 
$D4$-$D8$ model. 
Using a sum rule similar to eq.~(\ref{eq:pionvmd}) in the case of pions above,
\begin{equation}\label{eq:mesonvmd}
\sum_{n=1}^{\infty} \frac{g_{v^n}}{M_{v^n}^2}g_{v^n v^m v^{\ell}} = \delta_{m \ell}, 
\end{equation}
one can show that all photon-meson-meson couplings are cancelled in our model. 
Because of this the photon interacts with a meson only through intermediate vector mesons, 
which is a realization of vector meson dominance (VMD) in electromagnetic scattering in this model.\\
Numerically, the convergence of the relevant integrals is even slower than in the pion case. However, we find
a vector meson dominance pattern that is very similar to the one described in \cite{Grigoryan:2007vg}
in the AdS/QCD hard wall model, where the dominant contributions to $F_{v^1}(q^2 =0)$ were shown to come 
from the first two bound states (to $10^{-3}$ accuracy). Here we find, e.g,
\begin{eqnarray}
F_{v^1}(q^2=0) &\approx& \sum_{n=1}^{6} \frac{g_{v^n}}{M_{v^n}^2}g_{v^n v^1 v^{1}} = 0.999863,\\
F_{v^2}(q^2=0) &\approx& \sum_{n=1}^{6} \frac{g_{v^n}}{M_{v^n}^2}g_{v^n v^2 v^{2}} = 0.999840.
\end{eqnarray}
In order to compute the various form factors we first obtain a general expression valid for the elastic 
and non-elastic cases. Then, for the elastic case, we calculate the electric, magnetic and  
quadrupole form factors and briefly touch upon the large $q^2$ behavior of the longitudinal, transverse and 
longitudinal-transverse form factors. Also, for the first vector and axial-vector excitations,  
$\rho(770)$ and $a_1(1260)$, we compute the electric radius and the magnetic and quadrupole moments. 
We conclude this section with a brief discussion of the so-called transition or non-elastic form factors. 
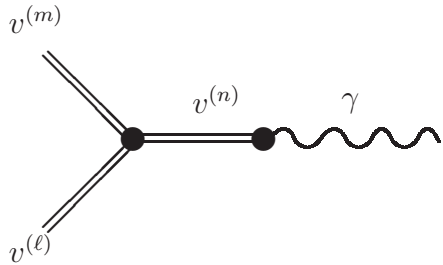
\begin{figure}[h]
\begin{center}
\vskip 3.cm
\begin{picture}(0,0)(5,0)
\setlength{\unitlength}{0.06in}
\rm
\thicklines 
%%%%%%%%%%%%%%%%%%%%%%%%%%%%%%%%%%%%%%%%% Vector meson %%%%%%%%%%%%%%%%%%%%%%
\put(-12,17){$v^{(m)}$}
\put(-9,15){\line(1,-1){7.5}}
\put(-8.6,15.4){\line(1,-1){7.5}} 
\put(-12,-3){$v^{(\ell)}$}
\put(-9,0){\line(1,1){7.5}}
\put(-8.6,-0.4){\line(1,1){7.5}}
\put(10.2,7.7){\circle*{2}}
\put(-1.2,8.1){\line(1,0){10.5}}
\put(-1.2,7.5){\line(1,0){10.5}}
\put(-1.2,7.7){\circle*{2}}
\put(4,10){$v^{(n)}$}
%%%%%%%%%%%%%%%%%%%%%%%%%%%%%%%%%%%%%%%%%%%%%%%%%  Photon  %%%%%%%%%%%%%%%%%%%%%
\put(17,10.5){$\gamma$}
\bezier{300}(10.5,7.5)(12.2,9.7)(13,7.5)
\bezier{300}(13,7.5)(14,6.5)(15,7.5)
\bezier{300}(15,7.5)(16.5,9.7)(17.5,7.5)
\bezier{300}(17.5,7.5)(18.5,6.5)(19.5,7.5)
\bezier{300}(19.5,7.5)(20.5,9.7)(21.5,7.5)
\bezier{300}(21.5,7.5)(22.5,6.5)(23.5,7.5)
\bezier{300}(23.5,7.5)(24.5,9.7)(25.5,7.5)
\end{picture}
\vskip 1.cm
\parbox{4.1 in}{\caption{Feynman diagram for vector meson form factor. A similar diagram holds for the axial-vector meson replacing the external lines $v^{(m)}, v^{(\ell)}$ by $a^{(m)}, a^{(\ell)}$. This graph also exhibits vector meson dominance.}}
\end{center}
%\label{fig:mesonff}
\end{figure}

\begin{figure}[h]
 \epsfig{file=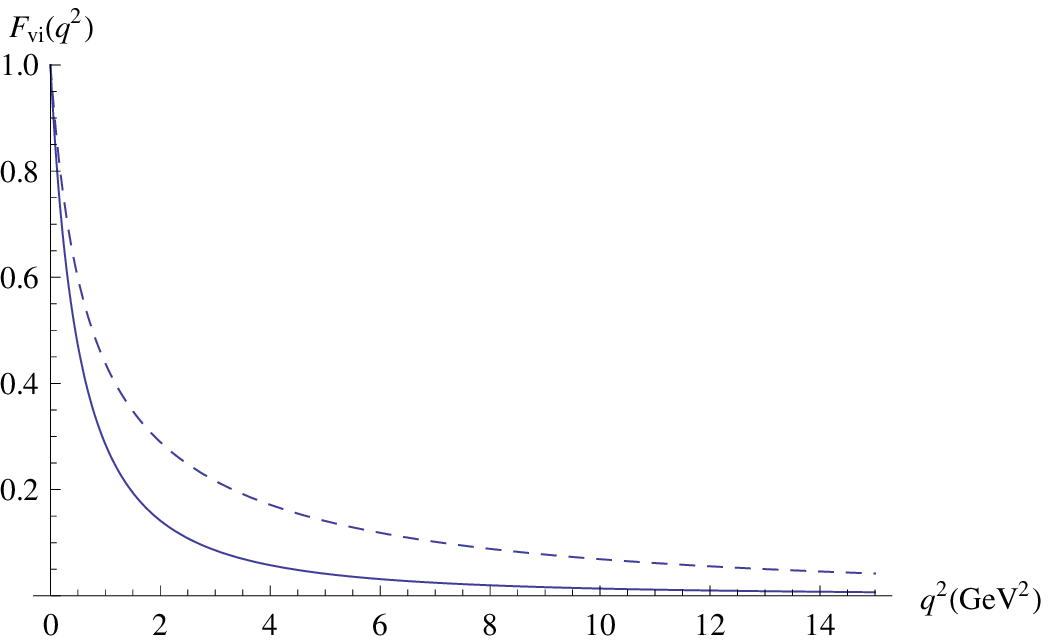, width=8cm}
 \epsfig{file=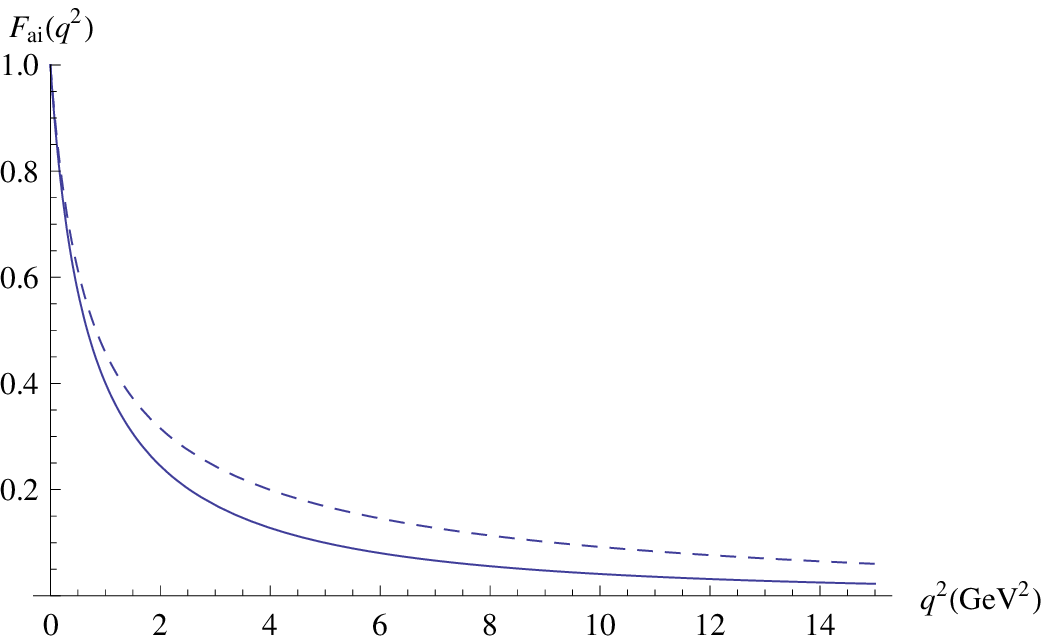, width=8cm}
\caption{Elastic form factors for for vector mesons (left) and axial-vector mesons (right) for the first two excited states: $i=1$ (solid line), $i=2$ (dashed line).}
\label{fig:mff1graphs}
\end{figure}

As before in the pion case, the form factors are calculated from the matrix elements of the electromagnetic current. 
The interaction of a vector meson with an off-shell photon is given by the matrix element
\begin{equation}
\langle v^{(m)\,a}(p), \epsilon \vert {\tilde J}^{\mu c}(q) \vert v^{(\ell) \,b}(p'), \epsilon' \rangle 
=  (2\pi)^4 \delta^4(p'-p-q) \, \langle v^{(m)\,a}(p), \epsilon \vert J^{\mu c}(0) \vert v^{(\ell) \,b}(p'), \epsilon' \rangle,
\end{equation}
where $v^{(m)}$ and  $v^{(\ell)}$ are the initial and final vector meson states with momenta $p$ and $p'=p+q$ and polarizations $\epsilon$ and $\epsilon'$. The operator ${\tilde J}^\mu$ is the Fourier transform of the electromagnetic current ${J}^\mu (x)$. 
This matrix element can be calculated from the corresponding Feynman diagram shown in figure 5. From the effective Lagrangian 
(\ref{eq:Lag2}), together with the interaction terms  (\ref{eq:Lag3}), we find 
\begin{eqnarray}
\langle v^{(m)\,a}(p), \epsilon \vert J^{\mu c}(0) \vert v^{(\ell) \,b}(p'), \epsilon' \rangle 
&=& \epsilon^\nu {\epsilon'}^\rho f^{abc} 
\left[ \eta_{\sigma\nu}(q-p)_\rho + \eta_{\nu\rho}(2p+q)_\sigma -  \eta_{\rho\sigma}(p + 2q)_\nu \right] \cr \cr
&& \times 
\sum_{n=1}^\infty {g_{v^n}g_{v^mv^nv^\ell}} \left[ \frac{\eta^{\mu\sigma} + \frac{q^\mu q^\sigma}{M_{v^n}^2}}{q^2+M_{v^n}^2} \right] 
\end{eqnarray}
where $M_{v^n}$ is the mass of the vector meson $v^{(n)}$. 
Defining the generalized vector meson form factor as 
\begin{equation}\label{eq:formvn}
F_{v^mv^\ell}(q^2)=\sum_{n=1}^\infty
\frac{g_{v^n}g_{v^nv^mv^\ell}}{q^2+M_{v^n}^2},
\end{equation} 
and using the sum rule eq.~(\ref{eq:mesonvmd}), we find  
\begin{eqnarray}\label{eq:longform}
\langle v^{(m)\,a}(p), \epsilon \vert J^{\mu c}(0) \vert v^{(\ell) \,b}(p'), \epsilon' \rangle 
&=& \epsilon^\nu {\epsilon'}^\rho f^{abc} 
\left[ \eta_{\sigma\nu}(q-p)_\rho + \eta_{\nu\rho}(2p+q)_\sigma -  \eta_{\rho\sigma}(p + 2q)_\nu \right] \cr \cr
&& \times \left\{ \left( {\eta^{\mu\sigma} - \frac{q^\mu q^\sigma}{q^2}} \right) F_{v^mv^\ell}(q^2) 
+ \delta_{m\ell} \frac{q^\mu q^\sigma}{q^2} \right\}. 
\end{eqnarray}
Taking into account the transversality of the vector meson polarizations, 
$\epsilon \cdot p =0 = \epsilon' \cdot p' $, and noting that the term involving the factor $\delta_{m\ell}$  
does not contribute since, in the elastic case $m=\ell$, we have $2p\cdot q + q^2 =0$, we obtain
\begin{eqnarray}\label{eq:longform2}
&& \langle v^{(m)\,a}(p), \epsilon \vert J^{\mu c}(0) \vert v^{(\ell) \,b}(p'), \epsilon' \rangle \cr\cr
&& \qquad = \epsilon^\nu {\epsilon'}^\rho f^{abc} 
\left[ \eta_{\nu\rho}(2p+q)_\sigma  + 2(\eta_{\sigma\nu} q_\rho - \eta_{\rho\sigma}q_\nu) \right] 
 \left( {\eta^{\mu\sigma} - \frac{q^\mu q^\sigma}{q^2}} \right) F_{v^mv^\ell}(q^2) \,.
\end{eqnarray}
Note also that this  matrix element satisfies the tranversality condition 
$q_\mu \langle v^{(m)} \vert J^{\mu}(0) \vert v^{(\ell)} \rangle =0$. \\
For axial-vector mesons a similar calculation can be done to find the form factors from the matrix element
$\langle a^{(m)\,a}(p), \epsilon \vert J^{\mu c}(0) \vert a^{(\ell) \,b}(p'), \epsilon' \rangle$. 
In this case we should replace the external vector meson lines
by the axial-vector mesons $a^{(m)}$ and $a^{(\ell)}$ in the Feynman diagram of figure 5. The internal vector meson line $v^{(n)}$, 
remains unchanged due to vector meson dominance. This is ensured by the sum rule (compare to eq.~(\ref{eq:mesonvmd})):
\begin{equation}\label{eq:vmdaxial}
\sum_{n=1}^{\infty} \frac{g_{v^n}}{M_{v^n}^2}g_{v^n a^m a^{\ell}} = \delta_{m \ell}, 
\end{equation}
Again, we can test this sum rule numerically, adding up the first 6 terms of the corresponding sums (for $a^{(2)}$ we sum up to 7 in order to improve 
numerical accuracy),
\begin{eqnarray}
F_{a^1}(q^2=0) &\approx& \sum_{n=1}^{6} \frac{g_{v^n}}{M_{v^n}^2}g_{v^n a^1 a^1} = 0.999872,\\
F_{a^2}(q^2=0) &\approx& \sum_{n=1}^{7} \frac{g_{v^n}}{M_{v^n}^2}g_{v^n a^2 a^2} =0.999804.
\end{eqnarray}
Thus, the generalized axial-vector meson form factor is
\begin{equation}\label{eq:forman}
F_{a^ma^\ell}(q^2)=\sum_{n=1}^\infty
\frac{g_{v^n}g_{v^na^ma^\ell}}{q^2+M_{v^n}^2} \ .
\end{equation}
\noindent
{\bf Elastic case.}
In order to obtain the elastic form factor for vector mesons, we consider the previous calculation imposing that the initial and final vector meson states $v^{(m)}$ coincide. Then, from eq.~(\ref{eq:longform2}), we find 
\begin{eqnarray}
&& \langle v^{(m)\,a}(p), \epsilon \vert J^{\mu c}(0) \vert v^{(m) \,b}(p'), \epsilon' \rangle \cr\cr
&& \qquad = f^{abc} \left\{ (\epsilon \cdot \epsilon') (2p+q)^\mu  + 
2\left[ \epsilon^\mu (\epsilon'\cdot q) - {\epsilon'}^\mu (\epsilon \cdot q) \right] \right\}
F_{v^m}(q^2) \,, 
\label{eq:formelastic}
 \end{eqnarray}
where $F_{v^m}(q^2)$ is the elastic form factor:
\begin{equation}\label{eq:formvnelast}
F_{v^m}(q^2)=\sum_{n=1}^\infty
\frac{g_{v^n}g_{v^nv^mv^m}}{q^2+M_{v^n}^2} \ .
\end{equation} 
Similar relations hold for 
the axial-vector mesons, merely replacing $v^{(m)}$ by $a^{(m)}$. We calculated numerically these sums from $n=1$ to
$n=6$ using the results obtained for the masses and couplings. 
The mass scale is $M_{\ast}^2=2.332\, (\text{GeV})^2$, as explained above in section \ref{sec:mesons}.
We plot in figure \ref{fig:mff1graphs} the elastic form factors for the vector meson 
$\rho$(770) ($v^{(1)}$) and axial-vector meson $a_1$(1260) ($a^{(1)}$). Note that when $q^2\to 0$, 
the vector and axial-vector form factors go to one, due to the sum rules (\ref{eq:mesonvmd}) and (\ref{eq:vmdaxial}), respectively.

\begin{figure}[h]
\epsfig{file=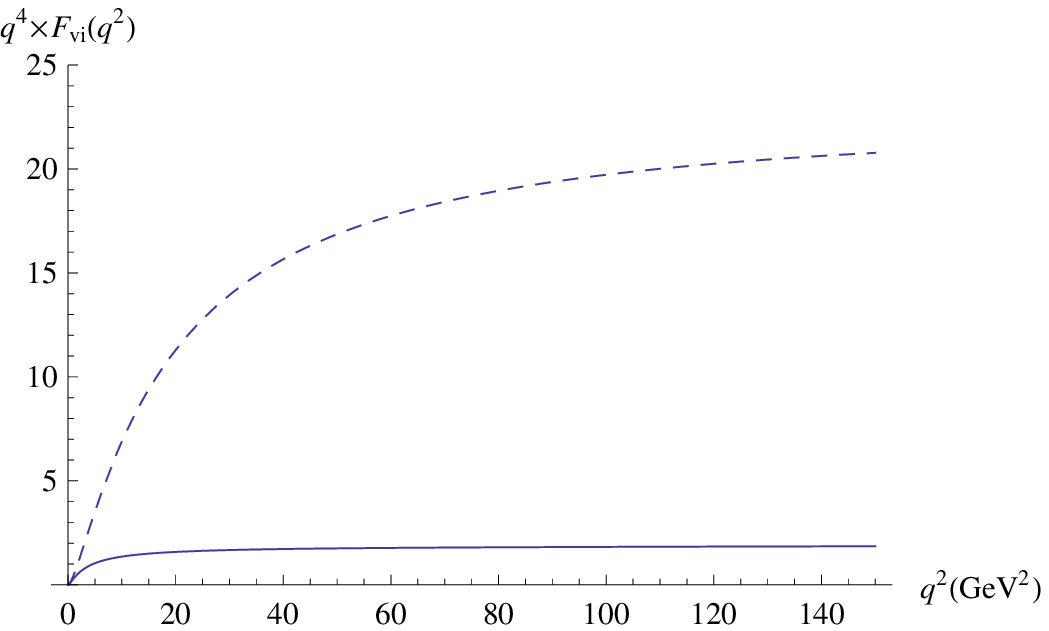, width=8cm}
\epsfig{file=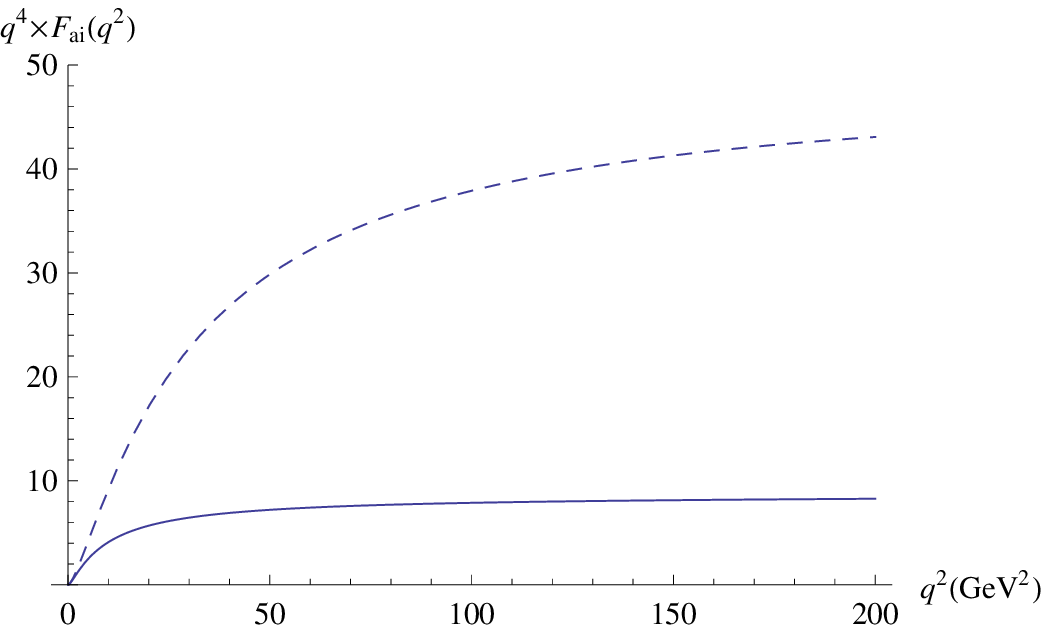, width=8cm}
\caption{$q^4$ times the elastic form factors for vector mesons (left) and axial-vector mesons (right) for the first two excited states: $i=1$ (solid line), $i=2$ (dashed line).}
\label{fig:mff2graphs}
\end{figure}

It is interesting to investigate the large $q^2$ behaviour of these elastic form factors.
Performing an expansion in powers of $q^{-2}$:  
\begin{eqnarray}
F_{v^m}(q^2) &=& \frac{1}{q^{2}}\sum_{n=1}^\infty g_{v^n}g_{v^nv^mv^m} \left( 1 - \frac{M_{v^n}^2}{q^2} + {\cal O}(q^{-4}) \right) \nonumber\\
F_{a^m}(q^2) &=& \frac{1}{q^{2}}\sum_{n=1}^\infty g_{v^n}g_{v^na^ma^m} \left( 1 - \frac{M_{v^n}^2}{q^2} + {\cal O}(q^{-4}) \right)  
\ ,  \label{eq:expelast}
\end{eqnarray} 
we see that the dominant terms would be of order $q^{-2}$. We calculated the coefficients of these terms  
using our numerical results with $n=1,...,6$ (or $7$, respectively). We found
\begin{eqnarray}
\sum_{n=1}^{6}
{g_{v^n}g_{v^nv^1v^1}} &\approx& 0.000176 ({\rm GeV})^2,\quad 
\sum_{n=1}^{6}{g_{v^n}g_{v^na^1a^1}} \approx 0.000220({\rm GeV})^2,
\cr
\sum_{n=1}^{6}
{g_{v^n}g_{v^nv^2v^2}} &\approx& -0.00309 ({\rm GeV})^2,\quad
\sum_{n=1}^{7}{g_{v^n}g_{v^na^2a^2}} \approx -0.00546 ({\rm GeV})^2,
\label{eq:superconv123}
\end{eqnarray}
These results indicate that the superconvergence relations 
\begin{equation}
\sum_{n=1}^\infty
{g_{v^n}g_{v^nv^mv^m}} = 0\; \; ,  \;\;
\sum_{n=1}^\infty
{g_{v^n}g_{v^na^ma^m}} = 0 \,,
\label{eq:superconv}
\end{equation}
hold in the KS model. Thus, from eqs.~(\ref{eq:expelast}) we expect that the form factors decrease 
approximately as $q^{-4}$ for large $q^2$. In figure \ref{fig:mff2graphs}, we plot  
the elastic form factors for the first and second vector and axial-vector states multiplied by $q^4$. 

\noindent
This corroborates the expected fall-off with $q^{-4}$. The deviation from the $q^{-4}$ dependence is related to the non-vanishing of the numerical sums (\ref{eq:superconv123}).

\noindent 
{\bf Electric, magnetic and quadrupole form factors.}
It is instructive to calculate the electric ($F_E$), magnetic ($F_M$) and quadrupole ($F_Q$) form factors from the previous results.
These form factors can be defined as (cf.~\cite{Grigoryan:2007vg})
\begin{equation}
F_E = F_1 + \frac{q^2}{6 p^2} \Big[ F_2 - ( 1 - \frac{q^2}{4 p^2})  F_3 \Big], \,\,\,
F_M = F_1 + F_2, \,\,\,
F_Q = - F_2 + \Big( 1 - \frac{q^2}{4p^2} \Big) F_3, 
\end{equation}
where $F_1 , F_2 $ and $F_3$ are related to the matrix element of the electromagnetic current for a spin one particle,   
\begin{eqnarray}
 \langle p, \epsilon \vert J^{\mu }(0) \vert p', \epsilon' \rangle 
&=&   (\epsilon \cdot \epsilon') (2p+q)^\mu F_1 ( q^2 )  + 
\left[ \epsilon^\mu (\epsilon'\cdot q) - {\epsilon'}^\mu (\epsilon \cdot q) \right] 
\left[ F_1 (q^2) + F_2(q^2) \right]  \nonumber\\
&+&  \frac{1}{p^2} ( q\cdot \epsilon') (q \cdot \epsilon ) (2p+q)^\mu F_3 (q^2) .
\label{eq:formdecomp}
\end{eqnarray}
Then, from eqs. (\ref{eq:formelastic}) and (\ref{eq:formdecomp}) we find that for a vector meson $v^{(m)}$,
\begin{equation}
F_1^{(v^m)} = F_2^{(v^m)} =  F_{v^m}\,,\,\,\;\;\;\; F_3^{(v^m)} = 0 \,,
\end{equation}
where $F_{v^m}$ is given by eq. (\ref{eq:formvnelast}). 
Therefore the electric, magnetic and quadrupole form factors predicted by the KS model 
for vector mesons are
\begin{equation}
F_E^{(v^m)} = ( 1  + \frac{q^2}{6 p^2} ) F_{v^m}
\,\,\, , \,\,\,
F_M^{(v^m)} = 2 F_{v^m} \,,
\,\,\,  \,\,\,
F_Q^{(v^m)} = - F_{v^m}  \,.
\end{equation}
The same formal results hold for the axial-vector mesons $a^{(n)}$.
In the following, we will estimate three important physical quantities from these form factors which are associated with 
the vector mesons: the electric radius, the magnetic and quadrupole moments. 

\begin{figure}[h]
 \epsfig{file=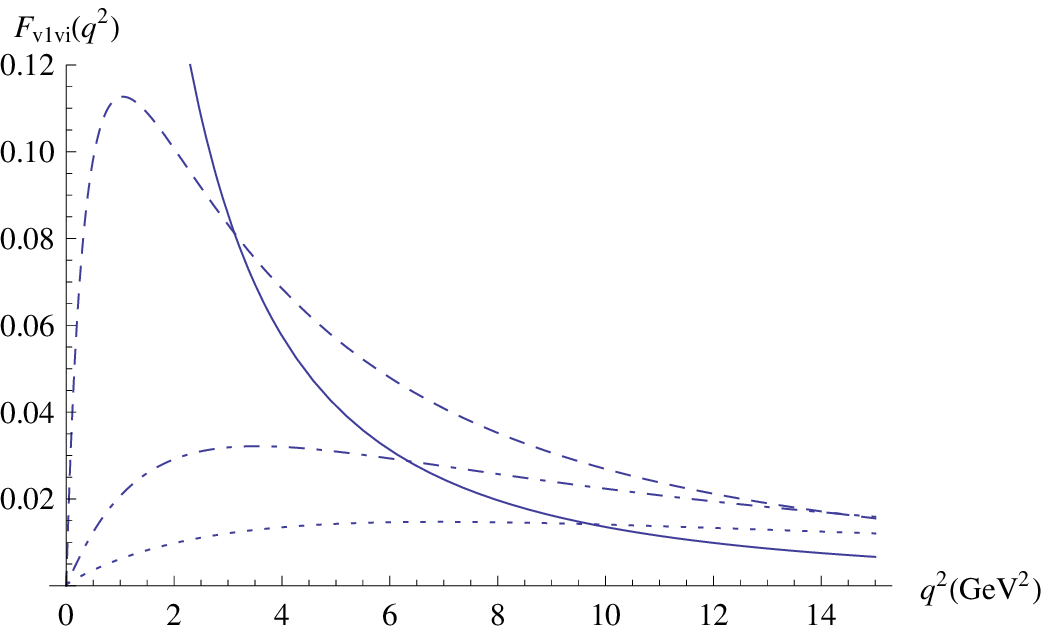, width=8cm}
 \epsfig{file=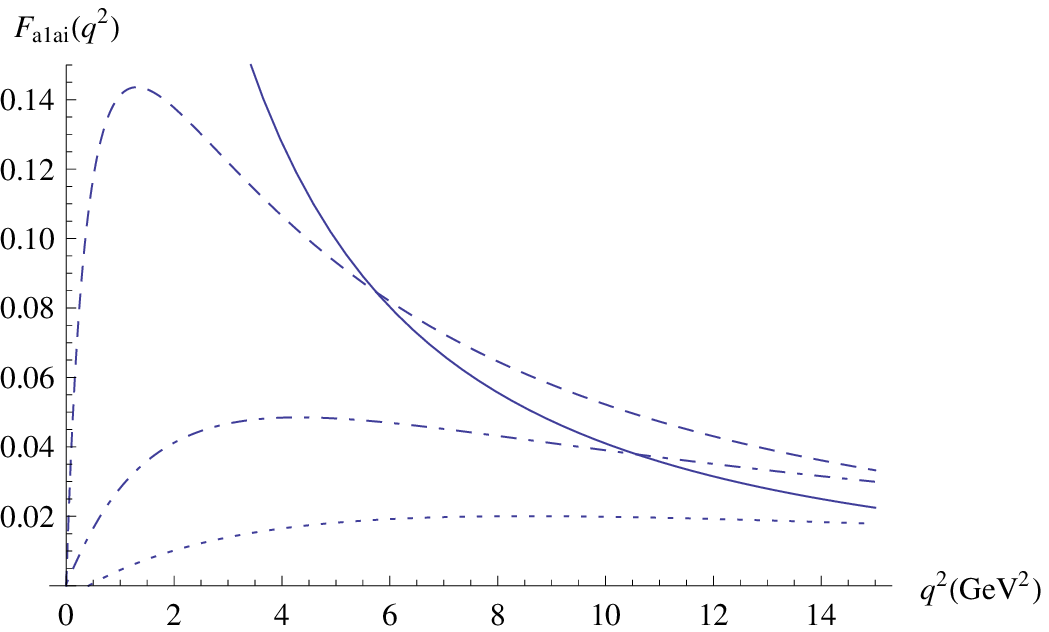, width=8cm}
\caption{Transition form factors for vector mesons $F_{v^1v^i}$ (left graphs) and axial-vector mesons $F_{a^1a^i}$ (right graphs), for $i=1$ (elastic case, solid line), $i=2$ (dashed line), $i=3$ (dot-dashed line) and $i=4$ (dotted line).}
\label{fig:mff3graphs}
\end{figure}

\noindent
{\bf Electric radius.}
The electric radius for the vector and axial-vector mesons are given by  
\begin{equation}
\langle r^2_{v^m}\rangle = -6\frac{\mathrm{d}}{\mathrm{d}q^2}F_E^{(v^m)}(q^2)|_{q^2=0}\;\; , \;\; 
\langle r^2_{a^m}\rangle = -6\frac{\mathrm{d}}{\mathrm{d}q^2}F_E^{(a^m)}(q^2)|_{q^2=0}\,.
\end{equation}
Using our numerical results for the form factors for the lowest excited states $\rho$ and $a_1$,
we find the electric radii (setting $p^2= M_{\rho}^2$ and $p^2= M_{a_1}^2$, respectively):
\begin{eqnarray}
 \langle r^2_{\rho}\rangle &=& \sum_{n=1}^{\infty} \frac{g_{v^n} g_{v^n v^1 v^1}}{M_{v^n}^4} -\frac{1}{M_{\rho}^2} \approx 0.3961 \,{\rm fm}^2, \\
 \langle r^2_{a_1}\rangle &=& \sum_{n=1}^{\infty} \frac{g_{v^n} g_{v^n a^1 a^1}}{M_{v^n}^4} -\frac{1}{M_{a_1}^2} \approx 0.3308 \,{\rm fm}^2.
\end{eqnarray}
Our results are slightly smaller than the results for the electric radii obtained in the $D4$-$D8$ model \cite{BallonBayona:2009ar} and the hard-wall model \cite{Grigoryan:2007vg}.\\
\noindent
{\bf Magnetic and quadrupole moments.}
Owing to the fact that $F_{v^n}$ and $F_{a^n}$ go to one as $q^2 \to 0$, the magnetic and quadrupole moments in the KS model formally reproduce the results obtained in other holographic models, e.g., the Sakai-Sugimoto model \cite{BallonBayona:2009ar} or the hard wall model \cite{Grigoryan:2007vg}:
\begin{eqnarray}
\mu &:=&  F_M(q^2)|_{q^2=0},\quad \quad \; \; \Rightarrow \mu_{v^n/a^n}=2 \\
D &:=& - \frac{1}{p^2} F_Q(q^2)|_{q^2=0}, \quad \Rightarrow D_{v^m} = - \frac{1}{M_{v^m}^2},\; D_{a^m} = - \frac{1}{M_{a^m}^2}.
\end{eqnarray}
Of course, the numerical values of these quantities in the KS model differ slightly from other holographic models, since the masses $M_{v^n}$ and $M_{a^n}$
differ from model to model.
\noindent 
{\bf Decomposition in terms of transverse and longitudinal polarizations.} 
It is instructive to study the form factors of vector mesons with specific polarizations.
The definition of the transverse and longitudinal polarizations and the corresponding form factors can be found in, e.g., \cite{Ioffe:1982qb,BallonBayona:2009ar}.
The large  $q^2$ behavior of these form factors in the KS model is determined
from $F_{v^m}$, eq. (\ref{eq:formvnelast}). 
We obtain the following asymptotic behaviors: $ F_{TT}^{(v^m)} \sim q^{-4} \,,\,
F_{LT}^{(v^m)} \sim q^{-3} \,,\, F_{LL}^{(v^m)} \sim q^{-2} \,$, in agreement with the $D4$-$D8$ model \cite{BallonBayona:2009ar} and QCD calculations (see, e.g., \cite{Ioffe:1982qb}).\\
\noindent
{\bf Non-elastic case.}
Finally, we consider the form factors for the case that the initial and final meson states are different. 
These are called generalized, transition or non-elastic form factors. 
The interaction with the photon is determined by 
the eqs.~(\ref{eq:formvn}), (\ref{eq:forman}), with $m \neq \ell$.
We calculate these form factors explicitly for the initial vector meson state $v^{(1)}$ and final states 
$v^{(m)}$ with $m=2,3,4$ and the same for axial-vector mesons. \\
The results are plotted in figure \ref{fig:mff3graphs}, compared with the corresponding elastic form factors.
Here, $q^2$ denotes the momentum transfer.

\begin{figure}[h]
 \epsfig{file=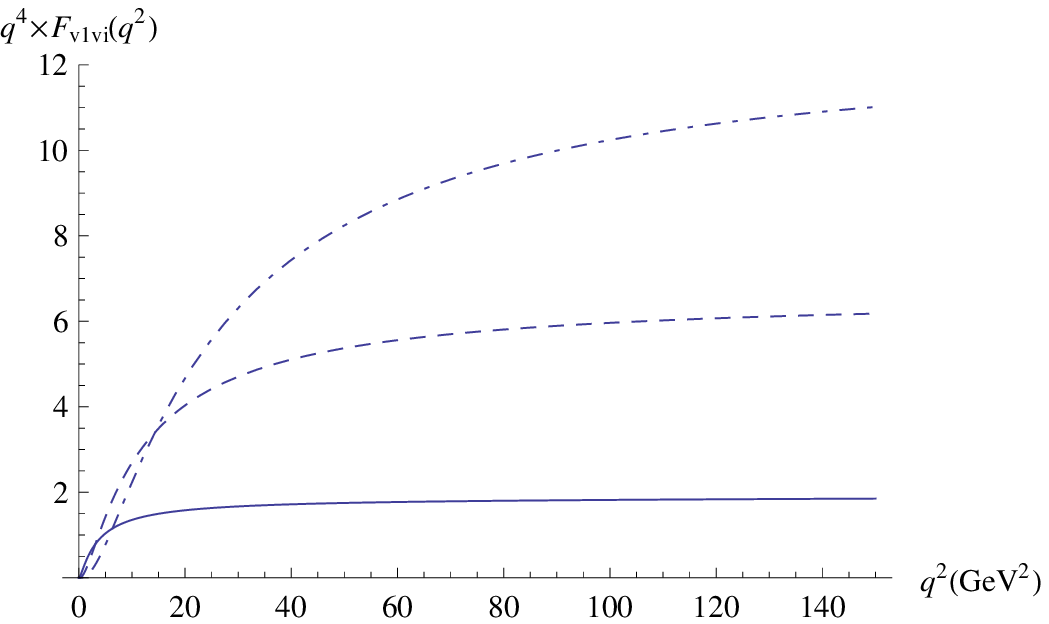, width=8cm}
 \epsfig{file=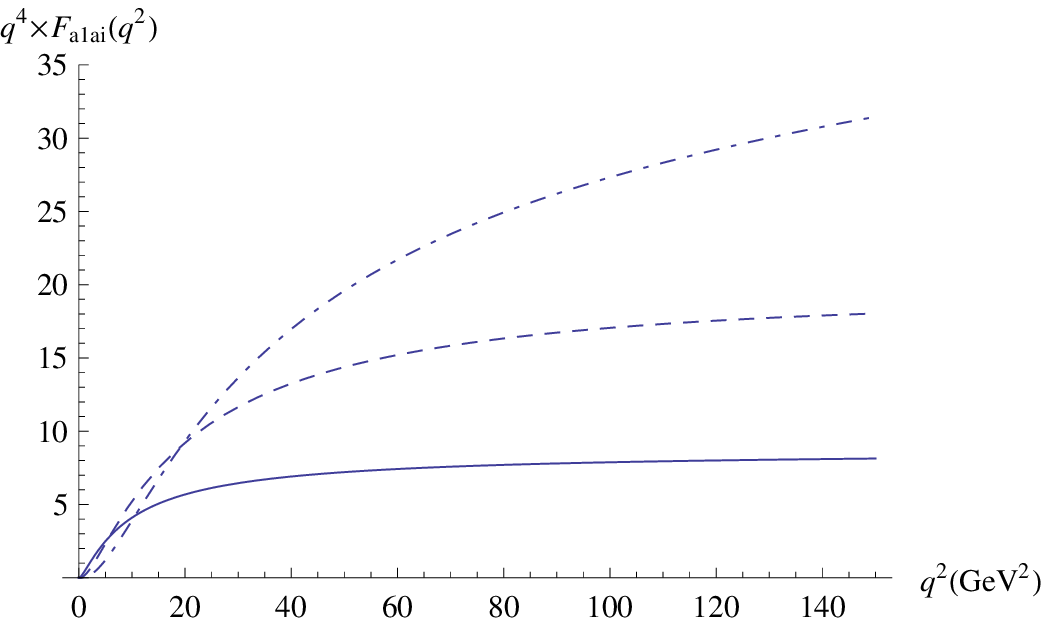, width=8cm}
\caption{$q^4$ times the transition form factors for vector mesons $F_{v^1v^i}$ (left graphs) and axial-vector mesons $F_{a^1a^i}$ (right graphs), for $i=1$ (elastic case, solid line), $i=2$ (dashed line), $i=3$ (dot-dashed line).}
\label{fig:mff4graphs}
\end{figure}
\noindent
Observe that, as $q^2\to 0$, the non-elastic form factors go to zero, 
while the elastic form factors approach one in this limit.
As we increase $q^2$, the form factors decrease at different rates. 
At small $q^2$, $F_{v^1}$ dominates, then $F_{v^1v^2}$ takes over, then $F_{v^1v^3}$, and so forth.
The same situation occurs for the axial-vector case. 
Physically, this can be explained by the fact that, as the momentum transfer increases,
the amplitude for producing heavier final states increases as well.\\ 
Regarding the large $q^2 $ dependence of these non-elastic form factors,  
we can make an expansion similar to eq. (\ref{eq:expelast}) but for different initial and final states. 
If the following superconvergence relations hold, 
 \begin{equation}
\sum_{n=1}^\infty
{g_{v^n}g_{v^nv^mv^\ell}} = 0 \; \; ,  \;\;
\sum_{n=1}^\infty
{g_{v^n}g_{v^na^ma^\ell}} = 0\,.
\label{eq:superconvinelast}
\end{equation}
we can again expect to find a $ q^{-4} $ behavior as in the elastic case. 
We calculated these sums from $n=1,..., 6$, for the relevant states, finding:
\begin{eqnarray}
    \sum_{n=1}^{6}
    {g_{v^n}g_{v^nv^1v^2}} &\approx& -0.000105 ({\rm GeV})^2
,\quad 
    \sum_{n=1}^{6}
    {g_{v^n}g_{v^na^1a^2}} \approx  -0.00316 ({\rm GeV})^2
,\cr 
    \sum_{n=1}^{6}
    {g_{v^n}g_{v^nv^1v^3}} &\approx& -0.00173 ({\rm GeV})^2
,\quad 
    \sum_{n=1}^{6}
    {g_{v^n}g_{v^na^1a^3}} \approx  0.0239({\rm GeV})^2
,\cr 
    \sum_{n=1}^{6}
    {g_{v^n}g_{v^nv^1v^4}} &\approx& 0.0155 ({\rm GeV})^2
,\quad 
    \sum_{n=1}^{6}
    {g_{v^n}g_{v^na^1a^4}} \approx  -0.185 ({\rm GeV})^2
 \,.    
\end{eqnarray}
These results indicate the validity of relations (\ref{eq:superconvinelast}). The numerical errors in $F_{v^1 v^i}$ and $F_{a^1a^i}$ increase as we 
increase $i$.
In order to check the large $q^2$ behavior, the transition form factors multiplied by $q^4$ are plotted in figure \ref{fig:mff4graphs}, along with the corresponding elastic form factors for comparison. 
We conclude that the transition form factors asymptotically approach the expected $q^{-4}$ dependence, within the numerical errors. 

\section{Conclusions}\label{sec:concl}
In this paper we studied the spectrum of vector and axial-vector mesons, along with the pion and meson form factors, in the holographic model of chiral symmetry breaking of Kuperstein and Sonnenschein. Moreover, we found that vector meson dominance is a feature of electromagnetic interactions of mesons in the KS model. We compared our numerical results to the Sakai-Sugimoto model as well as expectations from QCD. Our results for the form factors are generally in good qualitative agreement with QCD results. The pion form factor obtained in our model agrees well even quantitatively with the data provided by the Jefferson Lab Fpi2 collaboration.
This agreement is somewhat surprising, since our model is not dual to (large $N_c$) QCD, but rather a quiver gauge theory with gauge group $SU(N_c) \times SU(N_c)$. The meson physics is governed by open string excitations on the $D7$ flavor branes and depends on the details of the $D7$-brane embedding. This embedding is in many ways similar to the embedding of flavor branes in the Sakai-Sugimoto model. However, the asymptotic behavior of the geometries is quite different. The asymptotic conformal invariance of the Kuperstein-Sonnenschein model may explain why our results at large $q^2$ are close to QCD.  
Another notable difference between the Sakai-Sugimoto model and the KS model, concerning numerics, is that, in many cases, convergence of certain integrals and summations is much slower and harder to achieve numerically, due to the slow fall-off of the KS wave functions for large $\tilde{z}$. Nevertheless, our results, especially for the pion form factor, clearly demonstrate the viability and phenomenological potential of the Kuperstein-Sonnenschein model.
It would be very interesting to improve our understanding of the differences between the D-brane embeddings with respect to QCD phenomenology. For example, one could try to classify possible D-brane geometries and the resulting five-dimensional effective actions and investigate their usefulness for phenomenological questions. We leave this question to future research.\\
In a more recent paper, Dymarsky, Kuperstein and Sonnenschein (DKS) \cite{Dymarsky:2009cm} study a similar, yet more complicated, model of chiral symmetry breaking. This model is based on the Klebanov-Strassler background \cite{Klebanov:2000hb} which can be considered a non-conformal generalization of the Klebanov-Witten background. It features confinement at low energies and flows (via a Seiberg duality cascade) to a theory close to ${\mathcal N}=1$ super Yang-Mills in the IR. With the addition of fundamental flavor degrees of freedom this becomes a QCD-like theory. We are presently investigating its phenomenology which we will report in a follow-up paper \cite{Ihl:ta}.

\section*{Acknowledgements}
The authors are partially supported by CAPES and CNPq (Brazilian research agencies).


\begin{thebibliography}{19}

\bibitem{Maldacena:1997re}
  J.~M.~Maldacena,
  ``The large N limit of superconformal field theories and supergravity,''
  Adv.\ Theor.\ Math.\ Phys.\  {\bf 2}, 231 (1998)
  [Int.\ J.\ Theor.\ Phys.\  {\bf 38}, 1113 (1999)]
  [arXiv:hep-th/9711200].
  %%CITATION = IJTPB,38,1113;%%
\bibitem{Witten:1998qj}
  E.~Witten,
  ``Anti-de Sitter space and holography,''
  Adv.\ Theor.\ Math.\ Phys.\  {\bf 2}, 253 (1998)
  [arXiv:hep-th/9802150].
  %%CITATION = 00203,2,253;%%
\bibitem{Sakai:2004cn}
  T.~Sakai and S.~Sugimoto,
  ``Low energy hadron physics in holographic QCD,''
  Prog.\ Theor.\ Phys.\  {\bf 113}, 843 (2005)
  [arXiv:hep-th/0412141].
  %%CITATION = PTPKA,113,843;%%
\bibitem{Sakai:2005yt}
  T.~Sakai and S.~Sugimoto,
  ``More on a holographic dual of QCD,''
  Prog.\ Theor.\ Phys.\  {\bf 114}, 1083 (2005)
  [arXiv:hep-th/0507073].
  %%CITATION = PTPKA,114,1083;%%
\bibitem{Witten:1998zw}
  E.~Witten,
  ``Anti-de Sitter space, thermal phase transition, and confinement in  gauge
  theories,''
  Adv.\ Theor.\ Math.\ Phys.\  {\bf 2}, 505 (1998)
  [arXiv:hep-th/9803131].
  %%CITATION = 00203,2,505;%%
\bibitem{Karch:2002sh}
  A.~Karch and E.~Katz,
  ``Adding flavor to AdS/CFT,''
  JHEP {\bf 0206}, 043 (2002)
  [arXiv:hep-th/0205236].
  %%CITATION = JHEPA,0206,043;%%
\bibitem{Erdmenger:2007cm}
  J.~Erdmenger, N.~Evans, I.~Kirsch and E.~Threlfall,
  ``Mesons in Gauge/Gravity Duals - A Review,''
  Eur.\ Phys.\ J.\  A {\bf 35}, 81 (2008)
  [arXiv:0711.4467 [hep-th]].
  %%CITATION = EPHJA,A35,81;%%
\bibitem{BallonBayona:2009ar}
  C.~A.~Ballon Bayona, H.~Boschi-Filho, N.~R.~F.~Braga and M.~A.~C.~Torres,
  ``Form factors of vector and axial-vector mesons in holographic D4-D8
  model,''
  JHEP {\bf 1001}, 052 (2010)
  [arXiv:0911.0023 [hep-th]].
  %%CITATION = JHEPA,1001,052;%%
\bibitem{Hong:2004sa}
  S.~Hong, S.~Yoon and M.~J.~Strassler,
  ``On the couplings of vector mesons in AdS/QCD,''
  JHEP {\bf 0604}, 003 (2006)
  [arXiv:hep-th/0409118].
  %%CITATION = JHEPA,0604,003;%%
\bibitem{Grigoryan:2007vg}
  H.~R.~Grigoryan and A.~V.~Radyushkin,
  ``Form Factors and Wave Functions of Vector Mesons in Holographic QCD,''
  Phys.\ Lett.\  B {\bf 650}, 421 (2007)
  [arXiv:hep-ph/0703069].
  %%CITATION = PHLTA,B650,421;%%
\bibitem{Brodsky:2007hb}
  S.~J.~Brodsky and G.~F.~de Teramond,
  ``Light-Front Dynamics and AdS/QCD Correspondence: The Pion Form Factor in
  the Space- and Time-Like Regions,''
  Phys.\ Rev.\  D {\bf 77}, 056007 (2008)
  [arXiv:0707.3859 [hep-ph]].
  %%CITATION = PHRVA,D77,056007;%%
\bibitem{Aharony:2006da}
  O.~Aharony, J.~Sonnenschein and S.~Yankielowicz,
  ``A holographic model of deconfinement and chiral symmetry restoration,''
  Annals Phys.\  {\bf 322}, 1420 (2007)
  [arXiv:hep-th/0604161].
  %%CITATION = APNYA,322,1420;%%
\bibitem{Strominger:1995ac}
  A.~Strominger,
  ``Open p-branes,''
  Phys.\ Lett.\  B {\bf 383}, 44 (1996)
  [arXiv:hep-th/9512059].
  %%CITATION = PHLTA,B383,44;%%
\bibitem{Kuperstein:2008cq}
  S.~Kuperstein and J.~Sonnenschein,
  ``A New Holographic Model of Chiral Symmetry Breaking,''
  JHEP {\bf 0809}, 012 (2008)
  [arXiv:0807.2897 [hep-th]].
  %%CITATION = JHEPA,0809,012;%%
\bibitem{Klebanov:1998hh}
  I.~R.~Klebanov and E.~Witten,
  ``Superconformal field theory on threebranes at a Calabi-Yau  singularity,''
  Nucl.\ Phys.\  B {\bf 536}, 199 (1998)
  [arXiv:hep-th/9807080].
  %%CITATION = NUPHA,B536,199;%%
\bibitem{Ouyang:2003df}
  P.~Ouyang,
  ``Holomorphic D7-branes and flavored N = 1 gauge theories,''
  Nucl.\ Phys.\  B {\bf 699}, 207 (2004)
  [arXiv:hep-th/0311084].
  %%CITATION = NUPHA,B699,207;%%
\bibitem{Kuperstein:2004hy}
  S.~Kuperstein,
  ``Meson spectroscopy from holomorphic probes on the warped deformed
  conifold,''
  JHEP {\bf 0503}, 014 (2005)
  [arXiv:hep-th/0411097].
  %%CITATION = JHEPA,0503,014;%%
\bibitem{Levi:2005hh}
  T.~S.~Levi and P.~Ouyang,
  ``Mesons and Flavor on the Conifold,''
  Phys.\ Rev.\  D {\bf 76}, 105022 (2007)
  [arXiv:hep-th/0506021].
  %%CITATION = PHRVA,D76,105022;%%
\bibitem{BoschiFilho:2002ta}
  H.~Boschi-Filho and N.~R.~F.~Braga,
  ``QCD/String holographic mapping and glueball mass spectrum,''
  Eur.\ Phys.\ J.\  C {\bf 32}, 529 (2004)
  [arXiv:hep-th/0209080].
  %%CITATION = EPHJA,C32,529;%%
\bibitem{BoschiFilho:2002vd}
  H.~Boschi-Filho and N.~R.~F.~Braga,
  ``Gauge/string duality and scalar glueball mass ratios,''
  JHEP {\bf 0305}, 009 (2003)
  [arXiv:hep-th/0212207].
  %%CITATION = JHEPA,0305,009;%%
\bibitem{deTeramond:2005su}
  G.~F.~de Teramond and S.~J.~Brodsky,
  ``The hadronic spectrum of a holographic dual of QCD,''
  Phys.\ Rev.\ Lett.\  {\bf 94}, 201601 (2005)
  [arXiv:hep-th/0501022].
  %%CITATION = PRLTA,94,201601;%%
\bibitem{Erlich:2005qh}
  J.~Erlich, E.~Katz, D.~T.~Son and M.~A.~Stephanov,
  ``QCD and a Holographic Model of Hadrons,''
  Phys.\ Rev.\ Lett.\  {\bf 95}, 261602 (2005)
  [arXiv:hep-ph/0501128].
  %%CITATION = PRLTA,95,261602;%%
\bibitem{BoschiFilho:2005yh}
  H.~Boschi-Filho, N.~R.~F.~Braga and H.~L.~Carrion,
  ``Glueball Regge trajectories from gauge/string duality and the Pomeron,''
  Phys.\ Rev.\  D {\bf 73}, 047901 (2006)
  [arXiv:hep-th/0507063].
  %%CITATION = PHRVA,D73,047901;%%
\bibitem{Kruczenski:2003be}
  M.~Kruczenski, D.~Mateos, R.~C.~Myers and D.~J.~Winters,
  ``Meson spectroscopy in AdS/CFT with flavour,''
  JHEP {\bf 0307}, 049 (2003)
  [arXiv:hep-th/0304032].
  %%CITATION = JHEPA,0307,049;%%
\bibitem{Kirsch:2006he}
  I.~Kirsch,
  ``Spectroscopy of fermionic operators in AdS/CFT,''
  JHEP {\bf 0609}, 052 (2006)
  [arXiv:hep-th/0607205].
  %%CITATION = JHEPA,0609,052;%%
\bibitem{Skyrme:1961vq}
  T.~H.~R.~Skyrme,
  ``A Nonlinear field theory,''
  Proc.\ Roy.\ Soc.\ Lond.\  A {\bf 260}, 127 (1961).
  %%CITATION = PRSLA,A260,127;%%
\bibitem{Horn:2006tm}
  T.~Horn {\it et al.}  [Jefferson Lab F(pi)-2 Collaboration],
  ``Determination of the Charged Pion Form Factor at Q2=1.60 and 2.45
  (GeV/c)2,''
  Phys.\ Rev.\ Lett.\  {\bf 97}, 192001 (2006)
  [arXiv:nucl-ex/0607005].
  %%CITATION = PRLTA,97,192001;%%
\bibitem{Tadevosyan:2007yd}
  V.~Tadevosyan {\it et al.}  [Jefferson Lab F(pi) Collaboration],
  ``Determination of the pion charge form factor for Q2=0.60-1.60 GeV2,''
  Phys.\ Rev.\  C {\bf 75}, 055205 (2007)
  [arXiv:nucl-ex/0607007].
  %%CITATION = PHRVA,C75,055205;%%
\bibitem{Ioffe:1982qb}
  B.~L.~Ioffe and A.~V.~Smilga,
  ``Meson Widths And Form-Factors At Intermediate Momentum Transfer In
  Nonperturbative QCD,''
  Nucl.\ Phys.\  B {\bf 216}, 373 (1983).
  %%CITATION = NUPHA,B216,373;%%
\bibitem{Dymarsky:2009cm}
  A.~Dymarsky, S.~Kuperstein and J.~Sonnenschein,
  ``Chiral Symmetry Breaking with non-SUSY D7-branes in ISD backgrounds,''
  JHEP {\bf 0908}, 005 (2009)
  [arXiv:0904.0988 [hep-th]].
  %%CITATION = JHEPA,0908,005;%%
\bibitem{Klebanov:2000hb}
  I.~R.~Klebanov and M.~J.~Strassler,
  ``Supergravity and a confining gauge theory: Duality cascades and
  chiSB-resolution of naked singularities,''
  JHEP {\bf 0008}, 052 (2000)
  [arXiv:hep-th/0007191].
  %%CITATION = JHEPA,0008,052;%%
\bibitem{Ihl:ta}
  M.~Ihl, C.~A.~Ballon~Bayona, M.~A.~C.~Torres, H.~Boschi-Filho, 
  ``Phenomenology of the Dymarsky-Kuperstein-Sonnenschein model,''
  {\it work in progress.}
\bibitem{Minasian:1999tt}
  R.~Minasian and D.~Tsimpis,
  ``On the geometry of non-trivially embedded branes,''
  Nucl.\ Phys.\  B {\bf 572}, 499 (2000)
  [arXiv:hep-th/9911042].
  %%CITATION = NUPHA,B572,499;%%
\bibitem{Gimon:2002nr}
  E.~G.~Gimon, L.~A.~Pando Zayas, J.~Sonnenschein and M.~J.~Strassler,
  ``A soluble string theory of hadrons,''
  JHEP {\bf 0305}, 039 (2003)
  [arXiv:hep-th/0212061].
  %%CITATION = JHEPA,0305,039;%%
\bibitem{Evslin:2007ux}
  J.~Evslin and S.~Kuperstein,
  ``Trivializing and Orbifolding the Conifold's Base,''
  JHEP {\bf 0704}, 001 (2007)
  [arXiv:hep-th/0702041].
  %%CITATION = JHEPA,0704,001;%%
\bibitem{Krishnan:2008gx}
  C.~Krishnan and S.~Kuperstein,
  ``The Mesonic Branch of the Deformed Conifold,''
  JHEP {\bf 0805}, 072 (2008)
  [arXiv:0802.3674 [hep-th]].
  %%CITATION = JHEPA,0805,072;%%


\end{thebibliography}
\end{document}